\documentstyle[12pt]{article}
\newcommand{\dd}{\mbox{\rm d}}
\newcommand{\DD}{\mbox{\rm D}}
\newcommand{\nnn}{\noindent}
\newcommand{\p}{\partial}
\newcommand{\be}{\begin{equation}}
\newcommand{\ee}{\end{equation}}
\newcommand{\bi}{\bibitem}

\newcommand{\vs}{\vspace}

\begin{document}
\baselineskip .5cm
\rightline{IJS-TP/96-10}

\thispagestyle{empty}

\vs{1.5cm}

\begin{center}
{\large \bf The Dirac-Nambu-Goto $p$-Branes as Particular Solutions
to a Generalized, Unconstrained Theory}\footnote{
Work supported by the Slovenian Ministry of Science and Technology
under Contract J1-7455-0106-96}

\vs{6mm}

Matej Pav\v si\v c\footnote{Email: MATEJ.PAVSIC@IJS.SI}

Jo\v zef Stefan Institute, Jamova 39, SI-1000 Ljubljana, Slovenia

\vs{2cm}

ABSTRACT

\end{center}

\vs{.8cm}

The theory of the usual, constrained $p$-branes is embedded into a larger
theory in which there is no constraints. In the latter theory the Fock-
Schwinger proper time formalism is extended from point-particles
to membranes of arbitrary dimension. For this purpose the
tensor calculus in the infinite
dimensional membrane space ${\cal M}$ is developed and an action which is
covariant under reparametrizations in ${\cal M}$ is proposed.
The canonical and Hamiltonian formalism is
elaborated in detail. The quantization appears to be straightforward
and elegant. No problem with unitarity arises. The conventional $p$-brane
states are particular stationary
solutions to the functional Schr\" odinger equation which describes
the evolution of a membrane's state with respect to the invariant
evolution parameter $\tau$. A $\tau$ dependent solution which corresponds
to the wave packet of a null $p$-brane is found.
It is also shown that states of a lower
dimensional membrane can be considered as particular states of a
higher dimensional membrane.

\vs{1cm}

\nnn Short title:  {\it Generalized, unconstrained $p$-branes}

\vs{.7cm}

\nnn PACS: 1117 Theories of strings and other extended objects

\newpage
\setcounter{page}{1}

{\bf 1. Introduction}

The theory of relativistic membranes of arbitrary dimension ($p$-branes)
which include point particles ($p = 0$) and strings ($p = 1$) is very
elegant and is explored in great detail \cite{1},\cite{1a}.
Many interesting results are
obtained, including the possibility of unifying the fundamental
interactions. In view of such importance of the relativistic $p$-branes it is
desirable to obtain a broader look at the theory and find some
interrelation not known so far. Such a goal is attempted in the present
paper.

A motivation is to use, instead of the Dirac-Nambu-Goto action which is
practically unquantizable for a generic $p$-brane, another action which is
quantizable. There exists a very elegant approach by Schild \cite{Schild}
in which the string action contains the first power of the determinant $g$ of
the worldsheet metric (and not the square root). The equations of motion
are those of a Nambu-Goto string and at the same time the gauge is
automatically fixed so that $g$ is constant. The string analog of the
Hamilton-Jacobi formalism was elaborated by Eguchi \cite{Eguchi}
who found that the
world sheet area has the role of evolution parameter. The Schild action is
a straightforward generalization of the well known point-particle action
\cite{2} ${1 \over 2} \int {\dd} \tau \, ({{{\dd}X^{\mu}}\over
{\dd \tau}})^2$ which I shall call here the Stueckelberg action.

In recent years I have been exploring another possible generalization of the
Stueckelberg action to $p$-branes \cite{2aa}-\cite{4}.
The generalized $p$-brane action looks essentially like the
Stueckelberg action, except for the integration over the additional
parameters $\xi^a$ due to the extension of the object, and contains the
invariant evolution parameter $\tau$ and the background fields
$\Lambda$, $\Lambda^a$. It is {\it covariant} with
respect to reparametrizations of $\tau$, but it is not {\it invariant}
and thus there is no constraints. In Ref. \cite{4} the
equations of motion for the case $\Lambda^a = 0$ were derived and the
theory was quantized. The aim of the present
paper is to consider the more general case $\Lambda^a \neq 0$. 
A very interesting
relation involving expressions for $p$-brane constraints (Eq. \ref{D1})
is derived.  The canonical and Hamiltonian formalism is elaborated in
detail . Then we quantize the theory and show that particular
stationary states (with respect to $\tau$) correspond
to states of the conventional, constrained $p$-brane theory. We also obtain
that the wave functional of a $p$-dimensional membrane can represent, in a
limiting case, a ($p-1$)-dimensional membrane. Since this process can
be continued, it holds that states of a lower dimensional membrane are
contained among the states of a higher dimensional membrane. Finally, we
consider a simple solution which corresponds to the case of a null
membrane discussed by Schild \cite{Schild} and  Roshchupkin et al.
\cite{Zheltukhin}.)

\vs{1cm}

{\bf 2. The space of unconstrained membranes}

\vs{5mm}

The basic objects of the theory we are going to discuss in this paper are
$n$-dimensional, arbitrarily deformable and hence unconstrained, membranes
${\cal V}_n$ living in an $N$-dimensional space $V_N$.
The dimensions $n$
and $N$, as well as the corresponding signatures are left unspecified at this
stage. An unconstrained membrane ${\cal V}_n$ is represented by the
embedding functions $X^{\mu} (\xi^a)$, $\mu = 0,1,2,...,N-1$, where
$\xi^a$, $a = 0,1,2,...,n-1$, are local parameters (coordinates) on
${\cal V}_n$. The set of all possible membranes ${\cal V}_n$, with $n$ fixed,
forms an infinite dimensional space ${\cal M}$. A membrane ${\cal V}_n$
can be considered as a point in ${\cal M}$ parametrized by coordinates
$X^{\mu} (\xi^a) \equiv X^{\mu (\xi)}$
which bear a discrete index $\mu $ and $n$ continuous
indices $\xi^a$. To the discrete index $\mu$ we can ascribe arbitrary numbers:
instead of $\mu = 0,1,2,...,N-1$ we may set $\mu ' = 1,2,...,N$ or
$\mu ' = 2,5,3,1,...,$ etc. In general,
\be
    \mu' = f(\mu)
\label{A1}
\ee

\nnn where $f$ is a transformation. Analogously, a continuous index
$\xi^a$ can be given arbitrary continuous values. Instead of
$\xi^a$ we may take ${\xi'}^a$ which are functions of
$\xi^a$ :
\be
       {\xi'}^a = f^a (\xi)
\label{A2}
\ee

\nnn As far as we consider, respectively, $\mu$ and $\xi^a$ as a
discrete and a continuous index of coordinates $X^{\mu (\xi)}$
in the infinite dimensional space ${\cal M}$, reparametrization of
$\xi^a$ is analogous to renumbering of $\mu$. Both kinds of
transformations, (\ref{A1}) and (\ref{A2}), refer to the same point
of the space ${\cal M}$; they are {\it passive transformations}. For
instance, under the action of (\ref{A2}) we have
\be
     {X '}^{\mu} (\xi') = {X '}^{\mu} \left ( f(\xi) \right ) = X^{\mu} (\xi)
\label{A3}
\ee

\nnn which says that the same point ${\cal V}_n$ can be described
either by functions $X^{\mu} (\xi)$ or ${X'}^{\mu} (\xi)$ (where we may
write ${X '}^{\mu} (\xi)$ instead of ${X '}^{\mu} (\xi')$
since $\xi '$ is a running parameter and can be
renamed into $\xi$).

Then there exist also {\it the active transformations} which transform
one point of the space ${\cal M}$ into another. Given a parametrization
of $\xi^a$ and a numbering of $\mu$, a point ${\cal V}_n$ of
${\cal M}$ with coordinates $X^{\mu} (\xi)$ can be transformed into
another point ${\cal V '}_n$ with coordinates ${X '}^{\mu} (\xi)$. Parameters
$\xi^a$ are now considered as "body fixed", so that distinct
functions $X^{\mu} (\xi)$, ${X'}^{\mu} (\xi)$ represent distinct
points ${\cal V}_n$ , ${\cal V'}_n$ of ${\cal M}$. Physically, these are
distinct membranes which may be deformed one with respect to the
other. Such a membrane is {\it unconstrained}, since all coordinates
$X^{\mu} (\xi)$ are necessary for its description \cite{4a}-\cite{4}.
In order to
distinguish an unconstrained membrane ${\cal V}_n$ from the corresponding
mathematical manifold $V_n$, we use different symbols ${\cal V}_n$
and $V_n$.

It may happen, in particular, that two distinct membranes
${\cal V}_n$ and ${\cal V '}_n$ both
lie on the same mathematical surface $V_n$, and yet they are physically
distinct objects, represented by different points in ${\cal M}$.

The concept of an unconstrained membrane can be illustrated by imaging
a rubber sheet spanning a surface $V_2$. The sheet can be deformed
from one configuration (le me call it ${\cal V}_2$) into another
configuration ${\cal V'}_2$ in such a way that both configurations
${\cal V}_2$, ${\cal V'}_2$ are spanning the same surface $V_2$.
The configurations ${\cal V}_2$, ${\cal V'}_2$ are described by functions
$X^i (\xi^1, \xi^2)$, $X'^i (\xi^1, \xi^2)$ ($i = 1,2,3$), respectively.
The latter functions, from the mathematical point of view, both represent
the same surface $V_2$, and can be transformed one into the other by a
reparametrization of $\xi^1, \xi^2$. But from the physical point of view,
$X^i (\xi^1, \xi^2)$ and $X'^i (\xi^1, \xi^2)$ represent two different
configurations of the rubber sheet.

The reasoning presented in the last few paragraphs implies that,
since our membranes are assumed to be arbitrarily deformable,
different functions $X^{\mu} (\xi)$
can always represent physically different membranes. This
justifies use of the coordinates $X^\mu (\xi)$ for the
description of points in ${\cal M}$. Later, when we shall consider
membrane's dynamics, we shall admit $\tau$-dependence of coordinates
$X^\mu (\xi)$. In this section, all expressions refer to a fixed value of
$\tau$, therefore we omit it in the notation.

In analogy to the finite dimensional case we can introduce the distance
${\dd} {\ell}$ in our infinite dimensional space ${\cal M}$ :
\be
    {\dd} {\ell}^2 = \int {\dd} \xi \, {\dd} \zeta {\rho}_{\mu \nu}
    (\xi,\zeta) \,
    {\dd} X^{\mu} (\xi) \, {\dd} X^{\nu} (\zeta) = {\rho}_{\mu(\xi )
    \nu(\zeta)} \, 
    {\dd} X^{\mu (\xi)} \, {\dd} X^{\nu (\zeta)} =
    {\dd} X^{\mu (\xi)} {\dd} X_{\mu(\xi)}
\label{A4}
\ee

\nnn where ${\rho}_{\mu \nu} (\xi ,\zeta) = {\rho}_{\mu(\xi)\nu(\zeta)}$
is the metric in
${\cal M}$ . In eq.(\ref{A4}) we use a notation, similar to one that is
usually used when dealing with more evolved functional expressions
\cite{Bardakci}, \cite{DeWitt}. In order to distinguish
continuous indices from the discrete ones, the former are written
within brackets. When we write $\mu(\xi)$ as a subsrcipt or superscript
this denotes a pair of indices $\mu$ and $(\xi)$ (and not that $\mu$ is a
function of $\xi$). We also use the convention that summation is
performed over repeated indices (such as $a , b$) and integration over
repeated continuous indices (such as $(\xi), (\zeta)$). 

The tensor calculus in ${\cal M}$ \cite{4} is analogous to one in a finite
dimensional space. The differential of coordinates ${\dd} X^{\mu} (\xi)
= {\dd} X^{\mu(\xi)}$ is a vector in ${\cal M}$. The coordinates
$X^{\mu(\xi)}$ can be transformed into new coordinates
${X '}^{\mu(\xi)}$ which are functionals of $X^{\mu(\xi)}$ :
\be
       {X'}^{\mu (\xi)} = F^{\mu (\xi)} [X]
\label{A5}
\ee

\nnn The transformation (\ref{A5}) is very important. It says that if
functions $X^{\mu} (\xi)$ represent a membrane ${\cal V}_n$ then
any other functions $X'^{\mu} (\xi)$ obtained from $X^{\mu} (\xi)$
by a functional transformation also represent the same membrane
${\cal V}_n$. In particular, under reparametrization of $\xi^a$
the functions $X^{\mu} (\xi)$ change into new functions; a
reparametrization thus manifests itself as a special functional
transformation which belongs to a subclass of the general
functional transformations (\ref{A5}).

Under a general coordinate transformation (\ref{A5}) a generic
vector $A^{\mu(\xi)} \equiv A^{\mu} (\xi)$ transforms as \footnote{
A similar formalism, but for a specific type of the functional
transformations (\ref{A5}), namely the reparametrizations which
functionally depend on string coordinates, was developed by
Bardakci \cite{Bardakci} }
\be
      A^{\mu(\xi)} = {{\p {X'}^{\mu(\xi)}} \over {\p X^{\nu(\zeta)}}}
     A^{\nu (\zeta)} \equiv \int {\dd} \zeta 
     {{\delta {X'}^{\mu} (\xi)} \over {\delta X^{\nu} (\zeta)}} A^{\nu} (\zeta)
\label{A6}
\ee

\nnn Similar transformations hold for a covariant vector $A_{\mu(\xi)}$,
a tensor $B_{\mu(\xi)\nu(\zeta)}$, etc. Indices are
lowered and raised, respectively by ${\rho}_{\mu(\xi)\nu(\zeta)}$
and ${\rho}^{\mu(\xi)\nu(\zeta)}$, the latter being
the inverse metric tensor satisfying
\be
     {\rho}^{\mu(\xi) \alpha (\eta)} {\rho}_{\alpha (\eta) \nu (\zeta)} =
     {{\delta}^{\mu(\xi)}}_{\nu(\zeta)}
\label{A8}
\ee

Suitable choice of the metric - assuring the invariance of the line
element (\ref{A4}) under the transformations (\ref{A2}) and (\ref{A5}) -
is for instance
\be
       {\rho}_{\mu(\xi) \nu (\zeta)} = \sqrt{|f|} \, \alpha \, g_{\mu \nu}
       \delta (\xi - \zeta)
\label{A9}
\ee

\nnn where $f \equiv {\rm det} \, f_{ab}$ is the determinant of the
induced metric $f_{ab} \equiv {\p}_a X^{\alpha} {\p}_b X^{\beta} \, 
g_{\alpha \beta}$ on the sheet $V_n$, $g_{\mu \nu}$ is the metric
tensor of the embedding space $V_N$ and $\alpha$ an arbitrary function
of $\xi^a$. 

With the metric (\ref{A9}) the line element (\ref{A4}) becomes
\be
    {\dd} {\ell}^2 = \int  {\dd} \xi \sqrt{|f|} \, \alpha \, 
    g_{\mu \nu} \, {\dd} X^{\mu(\xi)} {\dd} X^{\nu(\xi)}
\label{A11}
\ee

\nnn Rewriting the abstract formulas back into
the usual notation with explicit integration we have
\begin{eqnarray}
    A^{\mu(\xi)} &=& A^{\mu} (\xi)  \\
    A_{\mu(\xi)} &=& \rho_{\mu (\xi) \nu (\zeta)} A^{\nu (\zeta)} =
    \int {\dd} \zeta \,  {\rho}_{\mu \nu} (\xi,\zeta)
    A^{\nu} (\xi) =
    \sqrt{|f|} \, \alpha \, g_{\mu \nu} A^{\nu} (\xi)
\label{A12}
\end{eqnarray}

\nnn The inverse metric is
\be
    {\rho}^{\mu(\xi)\nu(\zeta)} = {1 \over {\alpha \sqrt{|f|}}} \, \,
    g^{\mu \nu} \delta (\xi - \zeta)
\label{A13}
\ee

Indeed, from (\ref{A8}), (\ref{A9}) and (\ref{A13}) we obtain
\be
     {{\delta}^{\mu(\xi)}}_{\nu(\zeta)} = \int  {\dd} \eta \, g^{\mu 
     \sigma} g_{\nu \sigma}
     \, \delta (\xi - \eta) \delta (\zeta - \eta) =
     {{\delta}^{\mu}}_{\nu} \delta (\xi - \zeta)
\label{A14}
\ee

The invariant volume element (measure) of our membrane space
${\cal M}$ is \cite{2a}
\be
    {\cal D} X = {\left ( \mbox{\rm Det} \, {\rho}_{\mu \nu} (\xi,\zeta) 
    \right )}
    ^{1/2} \prod_{\xi,\mu} {\dd} X^{\mu} (\xi)
\label{A15}
\ee

\nnn Here Det denotes a continuum determinant taken over $\xi ,\zeta$
as well as over $\mu, \nu$. In the case of the diagonal metric (\ref{A9})
we have
\be
      {\cal D} X = \prod_{\xi ,\mu} \left ( \sqrt{|f|} \alpha 
       \, |g|\right ) ^{1/2}
       {\dd} X^{\mu} (\xi)
\label{A16}
\ee

\nnn As in a finite dimensional space we can now define covariant
derivative also in ${\cal M}$. For a scalar functional $A[X (\xi)]$
the covariant functional derivative coincides with the ordinary
functional derivative:
\be
    A_{;\mu(\xi)} = {{\delta A} \over {\delta X^{\mu} (\xi)}}
    \equiv A_{,\mu(\xi)}
\label{A16a}
\ee

\nnn But in general a geometric object in ${\cal M}$ is a tensor of
arbitrary rank, ${A^{{\mu}_1 (\xi_1) \mu_2 (\xi_2)...}}
_{\nu_1 (\zeta_1) \nu_2 (\zeta_2)...}$,
which is a functional of $X^{\mu} (\xi)$, and its covariant derivative
contains the affinity ${\Gamma}_{\nu(\zeta)\sigma(\eta)}^{\mu(\xi)}$
composed of the metric (\ref{A9}) \cite{4}. For instance, for a vector
we have
\be 
    {A^{{\mu}(\xi)}}_{; \nu(\zeta)} = {A^{\mu(\xi)}}_{, \nu(\zeta)}
    + {\Gamma}_{\nu(\zeta)\sigma(\eta)}^{\mu(\xi)}
    A^{\sigma(\eta)}
\label{A17}
\ee

Let the alternative notations for ordinary and covariant functional
derivative be ana\-logous to ones used in a finite dimensional space:
\be
     {{\delta } \over {\delta X^{\mu} (\xi)}} \equiv
     {{\p} \over {\p X^{\mu (\xi)}}} \equiv {\p}_{\mu(\xi)} 
\quad , \quad \quad {{\DD} \over {{\DD} X^{\mu} (\xi)}} \equiv
     {{\DD} \over {{\DD} X^{\mu (\xi)}}} \equiv {\DD}_{\mu(\xi)}
\label{A18}
\ee

\vs{1cm}
\newpage
{\bf 3. The classical theory of unconstrained membranes}

{\bf 3.1. The dynamics of unconstrained membranes}

\vs{5mm}

Now we are going to describe motion of a membrane ${\cal V}_n$ in an embedding
space $V_N$. For this purpose we introduce an extra parameter,
the evolution time $\tau$, and assume that the embedding functions
depend also on $\tau$, so that the parametric equation of a moving
membrane is
\be
     X^{\mu} = X^{\mu} (\tau, \xi^a)
\label{1}
\ee

We assume that the action is given by the following expression
\be
    I = {{\kappa} \over 2} \int {\dd} \tau \, {\dd}^n \xi \, \sqrt{|f|} \,
    \left ( {1 \over {\Lambda}} \, g_{\mu \nu} ({\dot X}^{\mu} +
    {\Lambda}^a \, \p_a X^{\mu} )({\dot X}^{\nu} + {\Lambda}^b \p_b X^{\nu})
    + \Lambda \right )
\label{2}
\ee

\nnn Here $\kappa$ is a constant, and
 dot denotes the derivative with respect to $\tau$.
If we treat $\Lambda$ and $\Lambda^a$ as the quantities
to be varied, i.e. the Lagrange multipliers, then one immediately finds
that they give the well known $p$-brane constraints \cite{7} and that the
action (\ref{2}) is equivalent \cite{5a} to the Dirac-Nambu-Goto action. In
Refs. \cite{2aa}-\cite{4}, on the contrary, we proposed to treat
$\Lambda$ and $\Lambda^a$ as the background, $\tau$-dependent fields
defined over membrane's coordinates
$\xi^a$. Eq.(\ref{2}) can then be considered as an approximation to a
full action with a kinetic term for $\Lambda$ and $\Lambda^a$. The latter
fields are presumably dynamical fields for a system of large number
of membranes, but are effectively the background fields for a single
member of the system. Though it would be very interesting to investigate
possible full actions and mechanisms leading to (\ref{2}) we prefer
first to explore the consequences of the action (\ref{2}). And if they
manifest themselves sufficiently interesting, this will justify a search for
a more fundamental principle behind the action (\ref{2}).

In a given parametrization, $\Lambda$ and $\Lambda^a$ are fixed,
prescribed functions of $\tau$ and
$\xi^a$. Under a reparametrization $\xi^a \rightarrow {\xi '}^a$
the quantities $\Lambda$ and $\Lambda^a$ are assumed to transform,
respectively, as a scalar and a vector in the
$n$-dimensional space $V_n$ associated with the membrane ${\cal V}_n$,
and under reparametrizations of $\tau$ they are assumed to transform
according to $\Lambda' = ({\dd \tau}/{\dd} \tau') \Lambda$ and ${\Lambda'}^a =
({\dd \tau}/{\dd} \tau') \Lambda^a$.
The action (\ref{2}) then does not change its form; it is given by the same
expression (\ref{2}) in which unprimed quantities are replaced by primed ones:
\be
    I' [{X '}^{\mu} {\Lambda '}, {\Lambda '}^a ] = 
    I [{X '}^{\mu} {\Lambda '}, {\Lambda '}^a ] =
    I [{X }^{\mu} {\Lambda }, {\Lambda }^a ]
\label{4}
\ee

\nnn Though the action (\ref{2}) is covariant with respect to
reparametrizations of $\xi^a$ and $\tau$ it is not invariant,
since the transformed action contains the transformed
functions $\Lambda ' \, , \; {\Lambda '}^a$. The latter
are not the quantities to be varied (not Lagrange multipliers)
and, consequently, the transformed action
$I' [{X '}^{\mu}]$ is not the same functional of the dynamical variables
as is the original action $I [X^{\mu}]$.

In the action (\ref{2}) the dimensions $n$ and $N$, and the
signatures of the corresponding manifolds $V_n$ and $V_N$ are left
unspecified. So the action (\ref{2}) contains many possible
particular cases. Especially interesting are the following ones:

{\it Case 1}. The manifold $V_n$ belonging to an unconstrained
membrane ${\cal V}_n$ has the signature (+ - - - ...) and corresponds to
an {\it n-dimensional worldsheet} with one time-like and $n-1$ space-like
dimensions. The index of the worldsheet
coordinates assumes the values $a = 0,1,2,...,n-1$.

{\it Case 2}. The manifold $V_n$ belonging to our membrane
${\cal V}_n$ has the signature (- - - - ...) and corresponds to a
space-like {\it p-brane}; therefore we take $n = p$. The index of
the membrane's coordinates $\xi^a$ assumes the values $a = 1,2,...,p$.

Throughout the paper we shall use the single formalism and apply it,
when convenient, either to the {\it Case 1} or to the {\it Case 2}.

When the dimension of the manifold $V_n$ belonging to
${\cal V}_n$ is $n = p + 1$
and the signature is (+ - - - ...), i.e. when we consider {\it Case 1},
then the action (\ref{2}) reduces to the action of
the usual Dirac-Nambu-Goto $p$-dimensional membrane (shortly $p$-brane) as
follows. Putting $\Lambda^a = 0$ and ${\dot X}^{\mu} = 0$
Eq.(\ref{2}) becomes
\be
    I = {{\kappa} \over 2} \int \Lambda \, {\dd} \tau \, {\dd}^n \xi
    \, \sqrt{|f|}
\label{3}
\ee

\nnn which is equal, when $\Lambda$ is independent of $\xi^a$, to the
minimal surface action, apart from the integration over ${\dd} \tau$, which
brings an extra constant factor ${{\kappa} \over 2} \int \Lambda {\dd} \tau$.

More generally, when $\Lambda^a \neq 0$, we obtain the action
(\ref{3}) if in Eq.(\ref{2}) we put
${\dot X}^{\mu} + \Lambda^a \p_a X^{\mu} = 0$. Then the $N$-velocity
${\dot X}^{\mu}$ is tangent to the surface $X^{\mu} (\xi)$, and the
$n$-velocity in $V_n$ is equal to $-\Lambda^a$.

An advantage of considering a $p$-brane's worldsheet
as a particular solution to the
equations of motion derived from (\ref{2}) is in the fact that the action
(\ref{2}) implies no dynamical constraints among the variables
$X^{\mu}$ and the corresponding canonical momenta $p_{\mu} =
(\kappa \sqrt{|f|}/\Lambda)({\dot X}_{\mu} + {\Lambda}^a \p_a X_{\mu})$.

A question now arises about a physical meaning of the parameter $\tau$
\cite{2},\cite{2aa}-\cite{4}. In the case of a point particle a
parametrization
of $\tau$ can be chosen such that $\Lambda$ is a constant. Then $\tau$ is
proportional to the proper time, i.e. to length of the worldline, and by
suitably choosing the constant of motion, $\tau$ becomes equal to
proper time.

In a generic case of an $n$-dimensional membrane the action (\ref{2})
can be written in a compact form by using the tensor notation of
${\cal M}$ space (Sec.2.) and taking $\alpha = \kappa/\Lambda$ in
the metric (\ref{A9}):
\be
    I = {1 \over 2} \int {\dd} \tau \left ( {\dot X}^{\mu (\xi)}
     {\dot X}_{\mu (\xi)} + K \right )
\label{P1}
\ee

\nnn where
\be
    K \equiv \int {\dd}^n \xi \, \sqrt{|f|} \, \Lambda \kappa
\label{P2}
\ee

\nnn and where, for simplicity, we consider for the moment the
case $\Lambda^a = 0$ \footnote{Such a choice is especially distinctive,
because the action
(\ref{2}) then becomes invariant (not only covariant) under
reparametrizations of membrane's coordinates $\xi^a$. Since the action is
stil not invariant under reparametrizations of the evolution parameter
$\tau$, there is no dynamical constraints, and all the variables
$X^{\mu} (\xi)$ and the momenta $p_{\mu} (\xi)$ remain independent
(see also refs. \cite{2aa}-\cite{4}).}.
The action (\ref{P1}) describes dynamics of a "point particle" in ${\cal M}$.
It is invariant under (i) renumbering of $\mu$ (Eq.(\ref{A1})), (ii)
reparametrizations of $\xi^a$ (Eq.(\ref{A2})), (iii) general coordinate
transformations in the embedding spacetime $V_N$ (including Lorentz
transformations, when $V_N$ is flat), and (iv) the general coordinate
transformations (\ref{A5}) in ${\cal M}$ (including the transformations
(ii) and (iii) as special cases). 
When $\Lambda$ is a constant in $\tau$, then the quantity 
${\dot X}^{\mu (\xi)} {\dot X}_{\mu (\xi)} - K$ is a constant of motion, $C$.
So we have ${\dd} X^{\mu (\xi)} {\dd} X_{\mu (\xi)} = (K + C) {\dd}
\tau^2$, i.e., the line element of the trajectory in ${\cal M}$
("proper time") is proportional to $\tau$. We can choose parametrization
of $\tau$ and/or the constant of motion $C$ such that $K + C = 1$; then
$\tau$ is just equal to proper time in ${\cal M}$. Simplicity of the
action (\ref{P1}) and the direct relation between $\tau$ and the
proper time in ${\cal M}$ justifies our choice to consider this
alternative
unconstrained $p$-brane action, instead of the Schild action \cite{Schild}.

In flat embedding spacetime $(g_{\mu \nu} =
\eta_{\mu \nu})$ the equations of motion, derived from the action
(\ref{2}) are
\be
     {1 \over {\sqrt{|f|}}} \, {{\dd} \over {{\dd} \tau}} \left ( \sqrt{|f|}
      \, \p X_{\mu} \, {1 \over {\Lambda}} \right ) +
      {1 \over {\sqrt{|f|}}} \, \p_a \left ( {\sqrt{|f|} \, ( \p X_{\mu} \,
      {{\Lambda^a} \over {\Lambda}} + \p^a X_{\mu} \, \mu }) \right ) = 0
\label{5}
\ee

\nnn where
\be
        \p X^{\mu} \equiv {\dot X}^{\mu} + {\Lambda}^a \p_a X^{\mu}
\label{6}
\ee
\be
   \mu \equiv {1 \over 2} \left ( {{\p X_{\mu} \, \p X^{\mu}} \over {\Lambda}}
   + \Lambda \right )
\label{7}
\ee

Let us first observe that $ {1 \over {\sqrt{|f|}}} \, \p_a (\sqrt{|f|} \,
\p^a X_{\mu} ) = {\DD}_a {\DD}^a X_{\mu}$ , where
${\DD}_a$ denotes the covariant
derivative with respect to coordinates $\xi^a$. The induced metric on
$V_n$ is $\gamma_{ab} = \p_a X^{\mu} \p_b X_{\mu}$. Since the covariant
derivative of $\gamma_{ab}$ is zero, we have the following identity
\be
    \p_c X^{\mu} \, {\DD}_a {\DD}_b X_{\mu} = 0
\label{8}
\ee

\nnn Contracting(\ref{5}) by $\p^c X^{\mu}$ and using (\ref{8}) we find
\be
    \p_c \mu = 
    - \, {1 \over {\sqrt{|f|}}} \, {{\dd} \over {{\dd} \tau}} \left (\sqrt{|f|}
      \, \p X_{\mu} \, {1 \over {\Lambda}} \right ) \p_c X^{\mu} -
     {1 \over {\sqrt{|f|}}} \, \p_a \left ( \sqrt{|f|} \, \p X_{\mu} \, 
      {{\Lambda^a} \over {\Lambda}} \right ) \p_c X^{\mu}
\label{9}
\ee

\nnn Inserting (\ref{9}) into the original equations of motion (\ref{5}) we
obtain
\be
    {b^{\nu}}_{\mu} \, \left [ {1 \over {\sqrt{|f|}}} \, {{\dd} \over
    {{\dd} \tau}} \left (\sqrt{|f|}
      \, \p X_{\nu} \, {1 \over {\Lambda}} \right ) +
      {1 \over {\sqrt{|f|}}} \, \p_a \left ( \sqrt{|f|} \, \p X_{\nu} \,
      {{\Lambda^a} \over {\Lambda}} \right )  \right ] +
      \mu {\DD}_a {\DD}^a X_{\mu} = 0
\label{10}
\ee

\nnn where
\be
          {b^{\nu}}_{\mu} \equiv {\delta^{\nu}}_{\mu} - \p_a X^{\nu} \p^a X_{\mu}
\label{11}
\ee

\nnn From Eq.(\ref{10}) it follows that the equation of minimal surface
\be
     {\DD}_a {\DD}^a X_{\mu} = 0
\label{12}
\ee

\nnn is satisfied, provided that
\be
    {b^{\nu}}_{\mu} \left [ {{{\dd} p_{\nu}} \over {{\dd} \tau}} +
    \p_a (p_{\nu} \Lambda^a) \right ] = 0
\label{13}
\ee

\nnn Here we have denoted
\be
    p_{\nu} = {{\kappa \sqrt{|f|} \, \p X_{\nu}} \over {\Lambda}}
\label{14}
\ee

\nnn where $p_{\nu} = \p {\cal L}/ \p {\dot X}^{\nu}$ is the
canonical momentum density belonging to the action (\ref{2}).

In particular, Eq.(\ref{13}) is satisfied when
\be
      \p X_{\mu} = 0
\label{15}
\ee

\nnn This is just the case we had in deriving the action (\ref{3}) from
(\ref{2}). However, we must take into account also Eq.(\ref{9}). We see
that the latter equation can be satisfied together with (\ref{15}) only when
\be
         \p_a \Lambda = 0
\label{16}
\ee

\nnn This is consistent with the fact that the action (\ref{3}) describes a
minimal surface if $\Lambda$ is independent of $\xi^a$.

In general, when $\p_a \Lambda \neq 0$,  from Eqs.(\ref{9}),(\ref{14}) we have
\be
     {{\dd} \over {{\dd} \tau}} (p_{\mu} \, \p_c X^{\mu}) +
     \p_a (\Lambda^a \, p_{\mu} \p_c X^{\mu}) = {1 \over 2} \,
     \p_c \Lambda \, {{\sqrt{|f|}} \over {\kappa}}
     \left ( {{p^2} \over {|f|}} - \kappa^2 \right )
\label{D1}
\ee

\nnn If a particular solution $X^{\mu} (\tau, \xi^a)$ to the membrane's
equations of motion satisfies
\be
    p_{\mu} \p_c X^{\mu} = 0
\label{D2}
\ee

\nnn then, because of (\ref{D1}), it automatically satisfies
\be
           p^2 - |f|\, \kappa^2 = 0
\label{D3}
\ee

The latter equations (\ref{D2}),(\ref{D3}) are the well known constraints
for an $n$-dimensional membrane \cite{7}. In particular, if we consider {\it
Case 2} they are just constraints for a usual $p$-brane, and a solution
$X^{\mu} (\tau, \xi)$ satisfying (\ref{D2}) then describes
a minimal $p+1$ dimensional surface.

In general, for our unconstrained membrane
\be
    p_{\mu} \, \p_a X^{\mu} \neq 0
\label{D4}
\ee
    
\nnn and $X^{\mu} (\tau, \xi)$ does not describe a minimal surface.
The momentum $p_{\mu}$ has a nonvanishing tangent component,
which means that in general there is the intrinsic motion within the membrane,
i.e. different parts of the membrane move relative to each other.
Such unconstrained membranes are closely related to the wiggly
membranes \cite{6},\cite{4a},\cite{5a}.

In particular, when Eq.(\ref{D2}) is satisfied, the momentum $p_{\mu}$
is perpendicular (in Minkowski space $V_N$) to the membrane $X^{\mu}$. 
We can distinguish two cases:

\ (i) $\Lambda^a = 0$; then Eq.(\ref{D2}) becomes $\p X^{\mu} \, \p_a
X_{\mu} = {\dot X}^{\mu} \, \p_a X_{\mu} = 0$ which means that there
is no intrinsic motion within the membrane (no wiggleness).

(ii) $\Lambda^a \neq 0$; then Eq.(\ref{D2}) becomes $\p X^{\mu} \, \p_a
X_{\mu} = ({\dot X}^{\mu} + \Lambda_c \, \p^c X^{\mu}) 
\p_a X_{\mu} = {\dot X}^{\mu}
\p_a X_{\mu} + \Lambda_a = 0$ which implies that there is an intrinsic
motion within the membrane. Our membrane, though sweeping a
minimal surface, is thus a wiggly membrane and wiggleness is
determined by $\Lambda^a$.
     
\newpage

{\bf 2.3 The canonical formalism and the Hamiltonian}

\vs{5mm}

The action (\ref{2}) implies no constraints, therefore the canonical
and Hamiltonian formalism \cite{5} is straightforward. The canonical momentum
density $p_{\mu} (\tau, \xi)$ satisfying the following equal $\tau$
Poisson bracket relations
\be
      \lbrace X^{\mu} (\xi), \, p_{\nu} (\xi ') \rbrace
       = {\delta^{\mu}}_{\nu} \,
      \delta (\xi - \xi ')
\label{17}
\ee
\be
     \lbrace X^{\mu} (\xi), \, X^{\nu} (\xi ') \rbrace
      = 0 \; , \quad \quad
     \lbrace p_{\mu} (\xi), \, p_{\nu} (\xi ')\rbrace = 0
\label{18}
\ee

\nnn The Poisson bracket of two generic functionals $A[X^{\mu} (\xi), \,
p_{\nu} (\xi)]$ and $B[X^{\mu} (\xi), \, p_{\nu} (\xi)]$ is defined by
\be
     \lbrace A, B \rbrace = \int {\dd}^n \xi '' \, \left ( 
     {{\delta A} \over {X^{\alpha} (\xi '')}} 
     \, {{\delta B} \over {\delta p_{\alpha} (\xi '')}} -
     {{\delta B} \over {X^{\alpha} (\xi '')}}
     \, {{\delta A} \over {\delta p_{\alpha} (\xi '')}}
     \right )
\label{19}
\ee

Now we are going to derive the Hamiltonian belonging to the action (\ref{2}). For
this purpose we vary both the field quantities $X^{\mu} (\tau, \xi)$ and
the boundary $B$ of the integration region $R$. In general, for an action
with a nonsingular Lagrangian density ${\cal L} (X^{\mu}, {\dot X}^{\mu},
\p_a X^{\mu})$ (which implies no constraints)
\be
       I = \int {\dd} \tau \, {\dd}^n \xi \, {\cal L} (X^{\mu}, {\dot X}^{\mu},
           \p_a X^{\mu})
\label{20}
\ee

\nnn we have
\begin{eqnarray}
     {\bar \delta} I &=& \int_R {\dd} \tau \, {\dd}^n \xi \, \delta {\cal L} +
     \int_{R - R'}  {\dd} \tau \, {\dd}^n \xi \,  {\cal L} = 
     \int_R {\dd} \tau \, {\dd}^n \xi \, \delta {\cal L} + \int {\dd} \tau \,
     \int_B {\dd} \Sigma_a \, {\cal L} \, \delta \xi^a +
     \int {\dd}^n \xi \, {\cal L} \, \delta \tau
     {\Biggl\vert}_{\tau_1}^{\tau_2} \nonumber \\
      &=& \int {\dd} \tau \, {\dd}^n \xi \,
     \left [ \delta {\cal L} + \p_a ({\cal L} \, \delta \xi^a) +
     {{\p} \over {\p \tau}} ({\cal L} \, \delta \tau) \right ]
\label{21}
\end{eqnarray}
     
\nnn This gives
\begin{eqnarray}
     {\bar \delta} I &=& \int {\dd} \tau \, {\dd} \xi \, \Biggl \lbrace \left [
     {{\p {\cal L}} \over {\p X^{\mu}}} - {{\p} \over {\p \tau}} \left (
     {{\p {\cal L}} \over {\p {\dot X}^{\mu}}} \right ) -
     \p_a {{\p {\cal L}} \over {\p \p_a X^{\mu}}} \right ] 
     \delta X^{\mu} \nonumber \\
     &+&
     \p_a \left ({{\p {\cal L}} \over {\p \p_a X^{\mu}}} \delta X^{\mu}
     + {\cal L} \, \delta \xi^a \right ) +
     {{\p} \over {\p \tau}} \left ( {{\p {\cal L}} \over {\p {\dot X}^{\mu}}} 
     \, \delta X^{\mu} + {\cal L} \, \delta \tau \right )
     \Biggr \rbrace
\label{22}
\end{eqnarray}

\nnn If we assume that the equations of motion are satisfied, then from
(\ref{22}) we have
\be
   {\bar \delta} I = \int {\dd} \tau \, \oint {\dd} \Sigma_a \, ({p^a}_{\mu}
   \delta X^{\mu} + {\cal L} \, \delta \xi^a ) + \int {\dd}^n \xi \, (p_{\mu}
   \delta X^{\mu} + {\cal L} \, \delta \tau ) \Biggl\vert_{\tau_1}^{\tau_2}
\label{22a}
\ee

\nnn where
\be
     p_{\mu} =  {{\p {\cal L}} \over {\p {\dot X}^{\mu}}}  \quad ,
     \hspace{1.5cm} {p^a}_{\mu} = {{\p {\cal L}} \over {\p \p_a X^{\mu}}}
\label{22b}
\ee

In the above equations (\ref{22}),(\ref{22a}) the field variation is defined
at a fixed value of the arguments $\tau \, , \xi^a$ :
\be
    \delta X^{\mu} \equiv {X '}^{\mu} (\tau , \xi) - X^{\mu} (\tau , \xi)
\label{23}
\ee

\nnn Let us now introduce, as usually in the field theory, the total
variation
\be
     {\bar \delta} X^{\mu} = {X '}^{\mu} (\tau ', \xi ') - X^{\mu} (\tau, \xi)
     = \delta X^{\mu} + {\dot X}^{\mu} \, \delta \tau + \p_a X^{\mu}\,
     \delta \xi^a
\label{24}
\ee

\nnn Rewriting Eq.(\ref{22a}) in terms of ${\bar \delta} X^{\mu}$ we obtain
\be
   {\bar \delta} I = \int {\dd} \tau \, \oint {\dd} \Sigma_a \, ({p^a}_{\mu}
   {\bar \delta} X^{\mu} - {T^a}_b \, \delta \xi^b - \theta^a \, \delta
   \tau) + \int {\dd}^n \xi \, (p_{\mu}
   {\bar \delta} X^{\mu} - {\cal H}_a \, \delta \xi^a - {\cal H}  \delta \tau )
   \Biggl\vert_{\tau_1}^{\tau_2}
\label{25}
\ee

\nnn where
\begin{eqnarray}
     {T^a}_b &=& {p^a}_{\mu} \, \p_b X^{\mu} - {\cal L}\, {\delta^a}_b 
     \label{26} \\
     \theta^a &=& {p^a}_{\mu} \, {\dot X^{\mu}} \label{27} \\
     {\cal H}_a &=& p_{\mu} \p_a X^{\mu} \label{28} \\
     {\cal H} &=& p_{\mu} {\dot X}^{\mu} - {\cal L} \label{29}
\end{eqnarray}     

In the first term of Eq.(\ref{25}) we can take the boundary at infinity and 
assume that the quantities ${p^a}_{\mu}, \; {T^a}_b , \; \theta^a$ vanish
at infinity, so that the variation is simply
\be
   {\bar \delta} I = \int {\dd}^n \xi \, (p_{\mu}
   {\bar \delta} X^{\mu} - {\cal H}_a \, \delta \xi^a - {\cal H}  \delta \tau )
   \Biggl\vert_{\tau_1}^{\tau_2} = G(\tau_2) - G(\tau_1)
\label{30}
\ee

\nnn The quantity
\be
     G(\tau) = \int {\dd}^n \xi \, (p_{\mu}
   {\bar \delta} X^{\mu} - {\cal H}_a \, \delta \xi^a - {\cal H}  \delta \tau )
\label{31}
\ee

\nnn defined at a fixed value of $\tau$ is {\it the generator of the total
variation} ${\bar \delta}$.

With the aid of the generator $G$ we can calculate variation of an arbitrary
functional $A[\tau, \xi^a, X^{\mu} (\xi), \, p_{\mu} (\xi) ]$
according to the Poisson bracket relation
\be
    \delta A = \lbrace A, G \rbrace
\label{32}
\ee

\nnn The latter relation can be proven as follows. Taking $A = X^{\mu}$ ,
using Eqs.(\ref{24}) and assuming the validity of Eq.(\ref{32}) we have
\be
    \delta X^{\mu} = {\bar \delta} X^{\mu} - \p_a X^{\mu} \, \delta \xi^a -
    {\dot X}^{\mu} \, \delta \tau = 
    \lbrace X^{\mu}, G \rbrace = {{\delta G} \over
    {\delta p_{\mu}}}
\label{33}
\ee

\nnn From (\ref{31}) and (\ref{33}) we have (after comparing the terms and
taking into account that the variations $\delta \xi^a$, $\delta \tau$ are
arbitrary):
\be
    \p_a X^{\mu} = \lbrace X^{\mu}, \, H_a \rbrace
\label{34}
\ee
\be
    {\dot X}^{\mu} = \lbrace X^{\mu}, \, H \rbrace
\label{35}
\ee

\nnn where
\be
     H_a = \int {\dd}^n \xi \, {\cal H}_a \quad , \quad \quad \quad 
     H = \int {\dd}^n \xi \, {\cal H}
\label{35a}
\ee

From (\ref{28}) we have that $H_a$ is a functional of $p_{\mu} (\xi)$
and $\p_a X^{\mu} (\xi)$. In the Hamiltonian density ${\cal H}$ (Eq.
(\ref{29})) we replace, using $p_{\mu} = \p {\cal L}/ \p {\dot X}^{\mu}$,
the velocity ${\dot X}^{\mu}$ by the momentum $p_{\mu}$,
so that $H = \int {\dd}^n \xi \, {\cal H}$ is a functional of
$X^{\mu} (\xi)$ and $p_{\mu} (\xi)$.

On the other hand, if we take $A = p_{\mu}$ we obtain
\be
    \delta p_{\mu} = {\bar \delta} p_{\mu} - \p_a p_{\mu} \, \delta \xi^a
    - {\dot p}_{\mu} \delta \tau = 
    \lbrace p_{\mu} , G \rbrace = - {{\delta G} \over {\delta X^{\mu}}}
\label{36}
\ee

\nnn where we have defined, in analogy to Eq.(\ref{24}), the total variation
of the momentum
\be
     {\bar \delta } p_{\mu} = {p'}_{\mu} (\tau', \xi ') - p_{\mu} (\tau, \xi)
     = \delta p_{\mu} + \p_a p_{\mu} \, \delta \xi^a + {\dot p}_{\mu} \delta
     \tau
\label{36a}
\ee

\nnn From (\ref{36}) we have
\be {\bar \delta} p_{\mu} = \int {\dd}^n \xi' \lbrace p_{\mu} (\xi), \, p_{\nu}
    (\xi') \rbrace
     {\bar \delta} X^{\nu} (\xi') \label{37} = 0 \ee
\be  \p_a p_{\mu} = \lbrace p_{\mu} , \, H_a \rbrace \label{38} \ee
\be {\dot p}_{\mu} = \lbrace p_{\mu} , \, H \rbrace \label{39} \ee

\nnn Using Eq.(\ref{28}) we can verify that the relations (\ref{34}) and
(\ref{38}) are satisfied identically. Using Eq.(\ref{29}) we find
that Eqs.(\ref{35}), (\ref{39}) are
equivalent to the equations of motion. Thus we have verified that Eq.
(\ref{32}) is satisfied for the variables $X^{\mu} (\xi)$ and $p_{\mu} (\xi)$.
For a generic functional we have
\begin{eqnarray}
   \delta A & = & \int {\dd}^n \xi' \, \left ( {{\delta A} \over {\delta X^{\mu}
   (\xi')}} \, \delta X^{\mu} (\xi') + {{\delta A} \over {\delta p_{\mu}
   (\xi')}} \, \delta p^{\mu} (\xi') \right ) \nonumber \\
   & = & \int {\dd}^n \xi' \, \left ({{\delta A} \over {\delta X^{\mu}
   (\xi')}} {{\delta G} \over {\delta p_{\mu} (\xi')}} -
   {{\delta A} \over {\delta p_{\mu} (\xi')}} \, {{\delta G} \over
   {\delta X^{\mu} (\xi')}} \right ) = \lbrace A, \, G \rbrace
\label{40}
\end{eqnarray}

In the last step of Eq.(\ref{40}) we have replaced $\delta X^{\mu}$ and
$\delta p_{\mu}$ according to the relations (\ref{33}), (\ref{36}) and thus
obtained the Poisson bracket. This completes the proof of Eq.(\ref{32}).

The total derivative of an arbitrary functional
$A[\tau, \xi^a, X^{\mu} (\xi), p_{\mu} (\xi)]$ with respect to $\tau$
and $\xi^a$ is
\be
     {{{\dd} A} \over {{\dd} \tau}} = {{\p A} \over {\p \tau}} +
     \int {\dd}^n \xi' \left ( {{\delta A} \over {\delta X^{\mu} (\xi')}}
     {\dot X}^{\mu} (\xi') + {{\delta A} \over {\delta p_{\mu} (\xi')}}
     {\dot p}_{\mu} (\xi') \right )
\label{B1}
\ee

\be
     {{{\dd} A} \over {{\dd} \xi^a}} = {{\p A} \over {\p \xi^a}} +
     \int {\dd}^n \xi' \left ( {{\delta A} \over {\delta X^{\mu} (\xi')}}
     \p_a X^{\mu} (\xi') + {{\delta A} \over {\delta p_{\mu} (\xi')}}
     \p p_{\mu} (\xi') \right )
\label{B2}
\ee

Replacing ${\dot X}^{\mu}$, ${\dot p}_{\mu}$, $\p_a X^{\mu}$ and
$\p_a p_{\mu}$ by the corresponding expressions (\ref{35}),(\ref{39}),
(\ref{34}) and (\ref{38}) Eqs.(\ref{B1}),(\ref{B2}) become
\be
      {{{\dd} A} \over {{\dd} \tau}} = {{\p A} \over {\p \tau}} +
      \lbrace A, \, H \rbrace
\label{B3}
\ee

\be
    {{{\dd} A} \over {{\dd} \xi^a}} = {{\p A} \over {\p \xi^a}} +      
     \lbrace A, \, H_a \rbrace
\label{B4}
\ee
     
The generator (\ref{31}) can be considered as the variation of the
Hamilton-Jacobi functional $S$:
\be
    G(\tau) = {\bar \delta} S = \int {\dd}^n \xi \, p_{\mu} (\xi)
    {\bar \delta} X^{\mu} (\xi) - H_a \, \delta \xi^a - H \, \delta \tau
\label{43aa}
\ee

\nnn Eq.(\ref{31}) or (\ref{43aa}) can be written in a simpler form by
introducing the variation
\be
    {\bar \delta}_{\tau} X^{\mu} = \delta X^{\mu} + {\dot X}^{\mu}
    \delta \tau = {X'}^{\mu} (\tau', \xi) - X^{\mu} (\tau, \xi)
\label{43a1}
\ee

\nnn so that the total variation defined in Eq.(\ref{24}) can be expressed
in terms of ${\bar \delta}_{\tau} X^{\mu}$ according to
\be
     {\bar \delta} X^{\mu} = {\bar \delta}_{\tau} X^{\mu} +
     \p_a X^{\mu} \, \delta \xi^a
\label{43a2}
\ee

\nnn Then the term $H_a \, \delta \xi^a$ in Eq.(\ref{43aa}) can be absorbed
into the first term and we obtain
\be
    G(\tau) = {\bar \delta} S = \int {\dd}^n \xi \, p_{\mu} (\xi) \,
    {\bar \delta}_{\tau} X^{\mu} (\xi) - H\, \delta \tau =
    p_{\mu (\xi)} {\dd} X^{\mu (\xi)} - H {\dd} \tau
\label{43a3}
\ee

\nnn In the last step of Eq.(\ref{43a3}) we have taken into account the
convention of the summation over the discrete index $\mu$ and the
integration over the continuous index $\xi$; instead of ${\bar \delta}_{\tau}
X^{\mu} (\xi)$ we have used the symbol ${\dd} X^{\mu (\xi)}$ which
denotes the differential of coordinates $X^{\mu (\xi)}$, i.e. an infinitesimal
vector in the membrane space ${\cal M}$ (see Sec.2.1). The differential
${\dd} X^{\mu (\xi)}$ is taken at fixed values of the indices
$\mu$ and $\xi$, but at any value of the evolution parameter $\tau$
(see Eq.(\ref{43a1}).

Now let us consider the momentum $p_{\mu (\xi)}$ as {\it the canonical
field} defined over the families of the trajectories $X^{\mu} (\tau, \xi)
\equiv X^{\mu (\xi)} (\tau)$ which are solutions to the equations of motion
for the action (\ref{20}). The vector field $p_{\mu (\xi)}$ is tangent to
the trajectories $X^{\mu (\xi)} (\tau)$ and is, in general,
a function of position
$X^{\mu (\xi)}$ in the membrane space ${\cal M}$, i.e. a functional of
$X^{\mu} (\xi)$,
\be
    p_{\mu (\xi)} =  p_{\mu (\xi)} (X^{\mu (\xi)}) \equiv
    p_{\mu} (\xi) [X^{\mu} (\xi)]
\label{43a4}
\ee

\nnn For the canonical field the Poisson bracket (\ref{18}) becomes
\be
   \lbrace p_{\mu (\xi)},p_{\nu (\xi')}\rbrace = \p_{\nu (\xi')} p_{\mu (\xi)} -
    \p_{\mu (\xi)} p_{\nu (\xi')} = 0
\label{43a5}
\ee

\nnn which is the expression for curl in ${\cal M}$. In addition, one
may require
\be
    D_{\mu (\xi)} p^{\mu (\xi)} = 0
\label{43a5a}
\ee

\nnn A system of trajectories for which (\ref{43a5}),(\ref{43a5a}) holds is
called a {\it coherent system of trajectories} \cite{Synge}.    
Because of (\ref{43a5}) the integral
\be
    \int {\dd} S = \int_a^b  (p_{\mu (\xi)} \, {\dd} X^{\mu (\xi)} -
    H \, \dd \tau ) = S[\tau_b, \, X_b^{\mu (\xi)}; \, \tau_a, \,
    X_a^{\mu (\xi)}]
\label{43a6}
\ee

\nnn is independent of the path between the points $a$ and $b$ in ${\cal M}$.    
Hence, by fixing the initial point $a$, we find that $S$ is a unique
function(al) of position $X^{\mu (\xi)}$ in ${\cal M}$, and also a function
of $\tau$:
\be
            S = S(\tau, X^{\mu (\xi)})
\label{S1}
\ee

\nnn This is {\it the Hamilton-Jacobi functional}. The total differential
of $S$ is
\be
     {\dd} S = {{\p S} \over {\p \tau}} \, {\dd} \tau +
     {{\p S} \over {\p X^{\mu (\xi)}}} \, {\dd} X^{\mu (\xi)} =
     p_{\mu (\xi)} \, {\dd} X^{\mu (\xi)} - H \, {\dd} \tau
\label{S2}
\ee

\nnn from which we have
\be
       {{\p S} \over {\p X^{\mu (\xi)}}} = p_{\mu (\xi)}
\label{S3}       
\ee

\be
       - {{\p S} \over {\p \tau}} = H
\label{S4}
\ee       

\nnn Eq.(\ref{S3}) tells us how the momentum field $p_{\mu (\xi)}$ is
related to $S$. (By the way, in the usual notation Eq.(\ref{S3})
reads $\delta S/\delta X^{\mu} (\xi) = p_{\mu} (\xi)$.) Eq.(\ref{S4})
is the functional Hamilton-Jacobi equation. It is equivalent to
the equations of motion.

So far we were discussing the general formalism for an arbitrary action
of the form (\ref{20}). Now we return to our particular membrane's
action (\ref{2}). The corresponding Hamiltonian can be straightforwardly
derived from (\ref{29}), (\ref{35a}) by using (\ref{14}) and (\ref{6}).
We obtain
\be
     H = \int {\dd}^n \xi \, \left [ \sqrt{|f|} \, 
     {\Lambda \over {2 \kappa}} \left ( {{p^{\mu} p_{\mu}} \over {|f|}}
     - \kappa^2 \right ) - \Lambda^a \, \p_a X^{\mu} p_{\mu} \right ]
\label{44}
\ee

The latter expression can be cast into the compact tensor notation of the
membrane space ${\cal M}$ (Sec. 2.1). We introduce the metric in ${\cal M}$
according to Eqs.(\ref{A9}),(\ref{A13}) with $\alpha = \kappa/\Lambda$ so
that the scalar product in the Hamiltonian (\ref{44}) can be written as
\be
   \int {\dd}^n \xi \, {{\Lambda} \over {\sqrt{|f|} \kappa}} \,p^{\mu} p_{\mu}     
    = \rho^{\mu(\xi) \nu (\xi')} p_{\mu (\xi)} p_{\nu (\xi')} =
    \rho_{\mu(\xi) \nu (\xi')} p^{\mu (\xi)} p^{\nu (\xi')} =
    p_{\mu (\xi)} p^{\mu (\xi)}
\label{44a}
\ee

\nnn where  $p_{\mu (\xi)} = p_{\mu} (\xi)$ and $p^{\mu (\xi)} =
\rho^{\mu(\xi) \nu (\xi')} p_{\nu (\xi')} = 
{{\Lambda} \over {\sqrt{|f|} \kappa}} \, p^{\mu} (\xi)$. Eq. (\ref{44})
then becomes
\be
      H = {1 \over 2} (p^{\mu (\xi)} p_{\mu (\xi)} - K ) -
      \Lambda^a \p_a X^{\mu (\xi)} p_{\mu (\xi)}
\label{44b}
\ee

\nnn where $K$ is defined in Eq.(\ref{P2}).

Since our action, when $\Lambda$ and $\Lambda^a$ are independent of $\tau$,
is invariant with respect to translations of the evolution
parameter $\tau \rightarrow \tau ' = \tau + a$, it follows from the
Noether theorem that the Hamiltonian is a constant of motion.
This follows also directly from the relation (\ref{B3}) which, for
$A = H$, gives ${\dot H} = 0$.

Using the specific Hamiltonian (\ref{44}) in the Hamiltonian equation
(\ref{39}) we have explicitly
\be
    {\dot p}_{\mu} = - \, {{\delta H} \over {\delta X^{\mu}}} =
    - \, {{\p {\cal H}} \over {\p X^{\mu}}} + \p_a {{\p {\cal H}}
    \over {\p \p_a X^{\mu}}} = - \, \kappa \, \p_a (\sqrt{|f|} \, \mu \,
    \p^a X_{\mu} + \sqrt{|f|} \, {{\Lambda^a} \over {\Lambda}} \p X_{\mu} )
\label{45}
\ee

\nnn where $\p X_{\mu}$ and $\mu$ are given in Eqs. (\ref{6}) and (\ref{7}),
respectively. Eq.(\ref{45}) coincides with the equation of motion (\ref{5})
derived directly from variation of the action (\ref{2}).

Similarly, we can explicitly verify also Eq.(\ref{35}):
\be
     {\dot X}^{\mu} = {{\delta H} \over {\delta p_{\mu} (\xi)}} =
     \int {\dd}^n \xi' \, \left [ {\Lambda \over {\sqrt{|f|} \kappa}} \,
     p^{\mu} (\xi') - \Lambda^a \p_a X^{\mu} (\xi') \right ]
     \delta (\xi' - \xi) = {\Lambda \over {\sqrt{|f|} \kappa}} \,
     p^{\mu} - \Lambda^a \p_a X^{\mu}
\label{46}
\ee

\nnn Using (\ref{6}) and (\ref{14}) we see that Eq.(\ref{46}) is indeed
an identity.

\vs{1cm}

{\bf 4. The quantum theory of unconstrained membranes}

\vs{3mm}

{\bf 4.1. The commutation relations and the Heisenberg equations of motion}

Since there is no constraints among the dynamical variables $X^{\mu} (\xi)$
and the corresponding canonical momenta $p_{\mu} (\xi)$, quantization of
the theory is straightforward. The classical variables become operators
and the Poisson brackets are replaced by commutators, taken at equal $\tau$.
Instead of Eqs.(\ref{17}),(\ref{18}) we have
\be
    [X^{\mu} (\xi), \, p_{\nu} (\xi')] = i\, {\delta^{\mu}}_{\nu}\, 
    \delta (\xi - \xi')
\label{47}
\ee
\be
     [X^{\mu} (\xi), \, X^{\nu} (\xi')] = 0 \; \; , \quad \quad 
     [p_{\mu} (\xi), \, p_{\nu} (\xi')] = 0
\label{48}
\ee

\nnn The quantum analog of Eq.(\ref{32}) for an operator $A(\tau, \xi)$ is
\be
    \delta A = - \, i \, [A, \, G]
\label{49}
\ee

\nnn implying, in the case of $G(\tau)$ (Eq.(\ref{31}), now considered as an
operator,
\be
      \p_a A = - \, i \, [A, \, H_a]
\label{50}
\ee
\be
     {\dot A} = -\, i \, [A, \, H]
\label{51}
\ee

Eq.(\ref{51}) is the Heisenberg equation of motion for an operator $A$. In
particular we have
\be
    {\dot p}_{\mu}(\xi) = - \, i \, [p_{\mu}(\xi), \, H]
\label{52a}
\ee
\be
      {\dot X}^{\mu}(\xi) = - \, i \, [X^{\mu}(\xi), \, H]
\label{52b}
\ee

We may use the representation in which the operators $X^{\mu} (\xi)$ are
diagonal and
\be
         p_{\mu}(\xi) = - \, i \, \left ({{\delta} \over 
         {\delta X^{\mu} (\xi)}} + {{\delta F} \over {\delta X^{\mu} (\xi)}}
         \right )
\label{52c}
\ee

\nnn where $F$ is a suitable functional of $X^{\mu}$ (see Eq.(\ref{65})).
The extra term in $p_{\mu}(\xi)$ is introduced in order to take into
account the metric (\ref{A9}) of the membrane space ${\cal M}$ (see
also Sec. 4.2)
In such a representation it is straightforward to calculate the
useful commutators
\be
    [p_{\mu} (\xi), \, \sqrt{|f|}] = - \, i \, \, {{\delta \sqrt{|f(\xi')|}}
    \over {\delta X^{\mu} (\xi)}} = i \, \p_a  \left ( \sqrt{|f|} \,
    \p^a X_{\mu} \, \delta (\xi - \xi') \right )
\label{54a}
\ee
\be
      [p_{\mu} (\xi), \p_a X^{\nu} (\xi') ] = - \, i \, {{
      \delta \p_a X^{\nu} (\xi')} \over {\delta X^{\mu} (\xi)}} = -\, i \,
      {\delta_{\mu}}^{\nu} \, \p_a \, \delta (\xi - \xi')
\label{54b}
\ee

\nnn For the Hamiltonian (\ref{44})  the explicit form of Eqs
(\ref{52a})and (\ref{52b}) can be obtained straightforwardly
by using the commutation relations (\ref{47}), (\ref{48}):
\be
   {\dot p}_{\mu} (\xi) = - \, \p_a \left [ {{\Lambda} \over {2 \kappa}} \,
   \p^a X_{\mu} \, \sqrt{|f|} ( {{p^2} \over {|f|}} + \kappa^2) \right ]
\label{54}
\ee

\begin{eqnarray}
     {\dot X}^{\mu} (\xi) &=& - \, i \, \int {\dd} \xi' \, [ X^{\mu}
     (\xi), \, {{\Lambda}\over {2 \kappa}} \sqrt{|f|} \left (
     {{p^{\alpha} (\xi') p_{\alpha} (\xi')} \over {|f|}} - \kappa^2 
     \right ) - \Lambda^a \p_a X^{\alpha} (\xi') p_{\alpha} (\xi')] \nonumber \\
     &=& {\Lambda \over {\sqrt{|f|} \kappa}} \,
     p^{\mu} (\xi') - \Lambda^a \p_a X^{\mu}
\label{55}
\end{eqnarray}

\nnn We recognise that the operator equations (\ref{54}), (\ref{55})
have the same form as the classical equations of motion (\ref{45}).

\vs{1cm}

{\bf 4.2. The Schr\" odinger representation}

The above relations (\ref{47})-(\ref{52b}),(\ref{54}),(\ref{55}) are valid
regardless of representation. A possible representation is one in
which the basic states $|X^{\mu} (\xi) \rangle$ have definite
values $X^{\mu} (\xi)$ of the membrane's position operators\footnote{
When necessary, we use symbols with hat, in order to distinguish operators
from their eigenvalues.}
${\hat X}^{\mu} (\xi)$. An arbitrary state $|a \rangle$ can be expressed
as
\be
    |a \rangle = \int |X^{\mu} (\xi) \rangle {\cal D} X^{\mu} (\xi) \langle
    X^{\mu} (\xi) | a \rangle
\label{56}
\ee

\nnn where the measure ${\cal D} X^{\mu} (\xi)$ is given in Eq.(\ref{A16})
with $\alpha = \kappa/\Lambda$.

Now we shall write the equation of motion for the wave functional
$\psi \equiv \langle X^{\mu} (\xi)|a \rangle$. We
adopt the requirement that, in the classical limit, the wave functional
equation should reproduce the Hamilton-Jacobi equation (\ref{S4});
the supplementary equation (\ref{S3}) has to arise from
the corresponding quantum equation. For this aim we admit that $\psi$
evolves with the evolution parameter $\tau$ and is a
functional of $X^{\mu} (\xi)$ :
\be
        \psi = \psi [\tau, X^{\mu} (\xi)]
\label{63}
\ee

\nnn It is normalized according to
\be
     \int {\cal D} X \, \psi^* \psi = 1
\label{F1}
\ee

\nnn which is a straightforward extension of the corresponding relation
$\int {\dd}^4 x \psi^* \psi = 1$ for the unconstraint point particle
in Minkowski spacetime \cite{2}-\cite{4}. It is important
to stress again \cite{4a}-\cite{4} that, since (\ref{F1}) is
satisfied at any $\tau$, the evolution operator $U$ which brings
$\psi (\tau) \rightarrow \psi (\tau') = U \psi (\tau)$ is
{\it unitary}.

The following equations are assumed to be satisfied ($\rho =
{\rm Det} \, \rho_{\mu(\xi) \nu(\xi')}$):
\be
    - \, i \hbar \,{1\over {|\rho|^{1/4}}} 
    \p_{\mu (\xi)} (|\rho|^{1/4} \psi) = {\hat p}_{\mu (\xi)} \psi
\label{64a}
\ee
\be        
     i \hbar \, {1\over {|\rho|^{1/4}}}
     {{\p (|\rho|^{1/4} \psi)} \over {\p \tau}} = H \psi
\label{64c}
\ee

\nnn where\footnote{
Using the commutator (\ref{54b}) we find that $\Lambda^a [p_{\mu (\xi)},
\p_a X^{\mu (\xi)}] = \int {\dd} \xi \, {\dd} \xi' \, \Lambda^a \,
[p_{\mu} (\xi),
\p_a X^{\nu} (\xi')] \, {\delta_{\nu}}^{\mu} \, \delta(\xi - \xi') = 0$
so that the order of operators in the second term of Eq.(\ref{64c1})
does not matter}
\be
          H = {1 \over 2} ({\hat p}^{\mu (\xi)} {\hat p}_{\mu (\xi)}
          - K ) - \Lambda^a \p_a X^{\mu (\xi)} {\hat p}_{\mu (\xi)}
\label{64c1}
\ee         

\nnn When the metric $\rho_{\mu(\xi) \nu(\xi')}$ in ${\cal M}$ explicitely
depend on $\tau$ (which is the case when ${\dot \Lambda} \ne 0$) such a
modified $\tau$-derivative in Eq.(\ref{64c}) is required \cite{DeWitt1}
in order to assure conservation of probability, as expressed by the
$\tau$-invariance of the integral (\ref{F1}).

The momentum operator given by
\be
       {\hat p}_{\mu (\xi)} = -i \hbar \left (\p_{\mu (\xi)} +
       {1\over 2} \Gamma_{\mu(\xi) \nu(\xi')}^{\nu(\xi')} \right )
\label{65}
\ee

\nnn where ${1\over 2} \Gamma_{\mu(\xi) \nu(\xi')}^{\nu(\xi')} =
|\rho|^{-1/4} \, \p_{\mu(\xi)} |\rho|^{1/4}$, satisfies the commutation
relations (\ref{47}),(\ref{48}) and is Hermitian with respect to the
scalar product $\int {\cal D} X \, \psi^* p_{\mu(\xi)}\psi$ in ${\cal M}$.
   
The expression (\ref{64c1}) for the Hamilton operator
$H$ is obtained from the corresponding classical
expression (\ref{44b}) in which the quantities
$X^{\mu (\xi)}$, $p_{\mu (\xi)}$ are replaced by the operators
${\hat X}^{\mu (\xi)}$, ${\hat p}_{\mu (\xi)}$.
There is an ordering ambiguity in the definition of 
${\hat p}^{\mu (\xi)} {\hat p}_{\mu (\xi)}$. Following the convention
in a finite dimensional curved space \cite{DeWitt1}, we 
use the identity $|\rho|^{1/4} {\hat p}_{\mu(\xi)}|\rho|^{-1/4} =
- \, i \hbar \, \p_{\mu(\xi)}$ and define
    $${\hat p}^{\mu (\xi)} {\hat p}_{\mu (\xi)} \psi =
    |\rho|^{-1/2} |\rho|^{1/4} {\hat p}_{\mu(\xi)}|\rho|^{-1/4}
    |\rho|^{1/2} \rho^{\mu(\xi) \nu(\xi')} |\rho|^{1/4} {\hat p}_{\nu(\xi')}
    |\rho|^{-1/4} \psi$$
\be    
     = - |\rho|^{-1/2} \p_{\mu(\xi)}(|\rho|^{1/2}
    \rho^{\mu(\xi) \nu(\xi')} \p_{\nu(\xi')} \psi) =
    - {\rm D}_{\mu(\xi)} {\rm D}^{\mu(\xi)} \psi
\label{65a}
\ee
  
Let us derive the classical limit of equations (\ref{64a}),(\ref{64c}).
For this purpose we write
\be
    \psi = A[\tau, X^{\mu} (\xi)] \; \mbox{\rm exp} \left [
    {i \over {\hbar}} S[\tau, X^{\mu}(\xi) \right ]
\label{66}
\ee

\nnn with real $A$ and $S$.

Assuming (\ref{66}) and taking the limit $\hbar \rightarrow 0$, Eq.(\ref
{64a}) becomes
\be
     {\hat p}_{\mu (\xi)} \psi = \p_{\mu (\xi)} S \, \psi    
\label{67}
\ee

\nnn If we assume that in Eq.(\ref{66}) $A$ is a slowly varying and $S$
a quickly varying functional of $X^{\mu} (\xi)$ we find that 
$\p_{\mu (\xi)} S$ is the expectation value of the momentum operator
${\hat p}_{\mu (\xi)}$.

Let us insert (\ref{66}) into Eq.(\ref{64c}). Taking the limit
$\hbar \rightarrow 0$, and writing separately the real and imaginary
part of the equation, we obtain
\be
      - \, {{\p S} \over {\p \tau}} = {1 \over 2} ( \p_{\mu (\xi)} S \,
      \p^{\mu (\xi)} S - K) - \Lambda^a \p_a X^{\mu (\xi)} \, \p_{\mu (\xi)} S
\label{70}
\ee
\be
      {1\over {|\rho|^{1/2}}} 
      {{\p \ } \over {\p \tau}} (|\rho|^{1/2} A^2)
      + {\mbox{\rm D}}_{\mu (\xi)} [A^2
      (\p^{\mu (\xi)} S - \Lambda^a \, \p_a X^{\mu (\xi)}) ] = 0
\label{71}
\ee

\nnn Eq.(\ref{70}) is just the functional Hamilton-Jacobi equation
(\ref{S4}) of the classical theory. Eq.(\ref{71}) is the continuity
equation, where $\psi^* \psi = A^2 $ is {\it the probability density}
and
\be
     A^2 \, (\p^{\mu (\xi)} S - \Lambda^a \, \p_a X^{\mu (\xi)}) 
      = j^{\mu (\xi)}
\label{71a}
\ee

\nnn is {\it the probability current}. While the covariant components
$\p_{\mu (\xi)} S$ (Eq.\ref{14})) form a momentum vector $p_{\mu}$,
the contravariant components form a vector $\p X^{\mu}$ (Eq.(\ref{6})):
\be
   \p^{\mu (\xi)} S = \rho^{\mu (\xi) \nu (\xi')} \, \p_{\nu (\xi')} S =
   \int {\dd} \xi' {{\Lambda} \over {\kappa \sqrt{|f|}}} \, \eta^{\mu \nu} \,
   \delta (\xi - \xi') {{\delta S} \over {\delta X^{\nu} (\xi')}} =
   {{\Lambda} \over {\kappa \sqrt{|f|}}} \eta^{\mu \nu} 
   {{\delta S} \over {\delta X^{\nu} (\xi)}} = \p X^{\mu}
\label{72}
\ee

\nnn where we have taken $\delta S/\delta X^{\nu} (\xi) =
p_{\nu} (\xi) = {{\kappa \sqrt{|f|}} \over {\Lambda}} \, 
\p X_{\nu} (\xi)$ (see Eq.(\ref{14})), and raised the index by
$\eta^{\mu \nu}$, so that $\p X^{\mu} (\xi) = \eta^{\mu \nu}
\p X_{\nu} (\xi)$. So we have $\p^{\mu (\xi)} S - \Lambda^a \,
\p_a X^{\mu (\xi)} = {\dot X}^{\mu (\xi)}$, and the current (\ref{71a})
is proportional to the velocity, as it should be.

Since Eq.(\ref{64c}) gives the correct classical limit, it is consistent
and can be taken as the equation of motion for the wave functional 
$\psi$. We shall call (\ref{64c}) {\it the (functional) Schr\" odinger
equation}. In general, it admits the following {\it continuity
equation}:
\be
     {1\over {|\rho|^{1/2}}} {{\p \ } \over {\p \tau}} 
     (|\rho|^{1/2} \psi^* \psi) 
    + {\mbox{\rm D}}_{\mu (\xi)} j^{\mu (\xi)}
    = 0
\label{73}
\ee

\nnn where
\begin{eqnarray}
     j^{\mu (\xi)} & = & {1 \over 2} \psi^* ({\hat p}^{\mu (\xi)} - 
     \Lambda^a \p_a X^{\mu (\xi)}) \psi + \mbox{\rm h.c.} \nonumber \\
     & = & - \, {i \over 2} (\psi^* \, \p^{\mu (\xi)} \psi -
     \psi \, \p^{\mu (\xi)} \psi^*) - \Lambda^a \p_a X^{\mu (\xi)}
     \psi^* \psi
\label{74}
\end{eqnarray}

For exercise, we prove bellow that the probability current (\ref{74})
satisfies (\ref{73}). First we observe that
\be
      {{\delta \, {\p'}_a X^{\mu} (\xi')} \over {\delta X^{\nu} (\xi)}}
      = {\p '}_a \left ( {{\delta X^{\mu} (\xi')} \over {\delta
      X^{\nu} (\xi')}} \right ) = {\delta ^{\mu}}_{\nu} \, {\p '}_a
      \delta (\xi' - \xi) = - \, {\delta^{\mu}}_{\nu}
      \p_a \delta (\xi' - \xi)
\label{75}
\ee

\nnn Then we calculate
   $$\p_{\nu (\xi)} (\Lambda^a \p_a X^{\mu (\xi')} \psi^* \psi) =
   {{\delta } \over {\delta X^{\nu} (\xi)}} (\Lambda^a \p_a X^{\mu}
   (\xi') \psi^* \psi)$$
\be
     = - \, \Lambda^a {\delta^{\mu}}_{\nu} \, \p_a \delta (\xi' - \xi)
     \psi^* \psi + \Lambda^a \p_a X^{\mu} (\xi') \left ( 
     \psi \, {{\delta \psi^*} \over {\delta X^{\nu} (\xi)}} +
     \psi^* \, {{\delta \psi} \over {\delta X^{\nu} (\xi)}} \right )
\label{76}
\ee

\nnn Multiplying (\ref{76}) by ${\delta^{\mu}}_{\nu} \delta (\xi' - \xi)
{\dd} \xi' {\dd} \xi$ suming over $\mu$, $\nu$ and integrating over 
$\xi'$, $\xi$, we obtain
\begin{eqnarray}
    \p_{\mu (\xi)} (\Lambda^a \p_a X^{\mu (\xi)} \psi^* \psi ) = &-& \, N
    \int {\dd} \xi' \, {\dd} \xi \, \Lambda^a \, \delta (\xi' -
    \xi) \, \p_a \delta (\xi' - \xi) \psi^* \psi \nonumber \\
     &+& \int {\dd} \xi \,
    \Lambda^a \, \p_a X^{\mu} (\xi) \left ( 
    \psi \, {{\delta \psi^*} \over {\delta X^{\mu} (\xi)}} +
     \psi^* \, {{\delta \psi} \over {\delta X^{\mu} (\xi)}} \right )
     \nonumber \\ 
     &=& \Lambda^a \, \p_a X^{\mu (\xi)} (\psi \, \p_{\mu (\xi)} \psi^* +
    \psi^* \, \p_{\mu (\xi)} \psi)
\label{77}
\end{eqnarray}

\nnn In Eq.(\ref{77}) we have taken ${\delta^{\mu}}_{\nu} {\delta^{\nu}}_
{\mu} = N$ and $\int {\dd} \xi \, \Lambda^a \delta (\xi' -
\xi) \, \p_a \delta (\xi' - \xi) = 0$. Next we take into account
${\mbox{\rm D}}_{\mu (\xi)} (\Lambda^a \p_a X^{\mu (\xi)} \psi^* \psi )
= ({\mbox{\rm D}}_{\mu (\xi)} \, \p_a X^{\mu (\xi)}) \Lambda^a
\psi^* \psi + \p_a X^{\mu (\xi)}\, \p_{\mu (\xi)} (\Lambda^a \psi^* \psi)$
and
\be
   {\mbox{\rm D}}_{\mu (\xi)}  \p_a X^{\mu (\xi)} = 
   \p_{\mu (\xi)} \p_a X^{\mu (\xi)} +
   \Gamma_{\mu (\xi) \nu (\xi')}^{\mu (\xi)} \, {\p '}_a X^{\nu (\xi')}
\label{78}
\ee

\nnn From (\ref{74}),(\ref{77}) and (\ref{78}) we have
   $${\DD}_{\mu (\xi)} j^{\mu (\xi)} = - \, {i \over 2} (
   \psi^* {\DD}_{\mu (\xi)} {\DD}^{\mu (\xi)} \psi -
   \psi {\DD}_{\mu (\xi)} {\DD}^{\mu (\xi)} \psi^*)$$
\be   
   - \Lambda^a \p_a
   X^{\mu (\xi)} (\psi^* \p_{\mu (\xi)} \psi + \psi \p_{\mu (\xi)} \psi^* 
   + \Gamma_{\mu(\xi) \nu(\xi')}^{\nu(\xi')}\, \psi^* \psi)
\label{84}
\ee

\nnn Using the Schr\" odinger equation $i |\rho|^{-1/4}
\p (|\rho|^{1/4}\psi)/\p \tau = H \psi$ and
the complex conjugate equation $-i |\rho|^{-1/4}\p 
(|\rho|^{1/4}\psi^*)/\p \tau = H^* \psi^*$ where
$H$ is given in (\ref{64c1}) we obtain that the continuity equation
(\ref{73}) is indeed satisfied by the probability density $\psi^* \psi$
and the current (\ref{74}). For the Ansatz (\ref{66}) the current (\ref{74})
becomes equal to the expression (\ref{71a}) as it should.

We notice that the term with 
$\Gamma_{\mu(\xi) \nu(\xi')}^{\nu(\xi')}$ in Eq.(\ref{84})
is canceled by the same type of the term in $H$. The latter term in $H$
comes from the definition (\ref{65}) of the momentum operator in a (curved)
membrane space ${\cal M}$, whilst the analogous term in Eq.(\ref{84})
results from the covariant differentiation. The definition (\ref{65}) of
${\hat p}_{\mu(\xi)}$, which is an extension of the definition introduced
by DeWitt \cite{DeWitt1}for a finite dimensional curved space,
is thus shown to be consistent also with the conservation
of the current (\ref{74}).

\vs{1cm}

{\bf 4.3 The stationary Schr\" odinger equation for a membrane}

\vs{5mm}

Evolution of a generic membrane's state $\psi[\tau, X^{\mu}(\xi)]$
is given by the $\tau$-dependent functional Schr\" odinger equation
(\ref{64c}) and the Hamiltonian (\ref{64c1}). We are now going to
consider solutions which have the form
\be
    \psi[\tau, X^{\mu}(\xi)] = e^{-iE \tau} \phi[X^{\mu}(\xi)]
\label{85}
\ee

\nnn where $E$ is a constant. We shall call it {\it energy}, since it has
analogous role as energy in non relativistic quantum mechanics.
Considering the case ${\dot \Lambda} = 0$, ${\dot \Lambda}^a = 0$ and
inserting
the Ansatz (\ref{85}) into Eq.(\ref{64c}) we obtain
\be
     \left ( -\, {1 \over 2} {\DD}^{\mu (\xi)} {\DD}_{\mu (\xi)} +
     i\, \Lambda^a \p_a X^{\mu (\xi)} (\p_{\mu (\xi)} +
     {1\over 2} \Gamma_{\mu(\xi) \nu(\xi')}^{\nu(\xi')})
     - {1\over 2} K \right ) \phi = E \, \phi
\label{86}
\ee

\nnn So far membrane's dimension and signature were not specified. Let us
now consider {\it Case 2} of Sec 2.2.     
All the dimensions of our membrane have the same
signature, and the index $a$ of membranes' coordinates assumes
the values $a = 1,2,..., n = p$. Assuming a real $\phi$ Eq.(\ref{86})
becomes
\be 
   \left ( -\, {1 \over 2} {\DD}^{\mu (\xi)} {\DD}_{\mu (\xi)} -
     {1\over 2} K - E \right ) \phi = 0
\label{88}
\ee

\be
    \Lambda^a \p_a X^{\mu (\xi)} (\p_{\mu (\xi)} + 
     {1\over 2} \Gamma_{\mu(\xi) \nu(\xi')}^{\nu(\xi')}) \phi = 0
\label{89}
\ee

\nnn These are equations for a {\it stationary state}. They remind us of 
the well known $p$-brane equations \cite{7}.

In order to obtain from (\ref{88}),(\ref{89}) the conventional
$p$-brane equations we have to assume that Eqs.(\ref{88}),(\ref{89})
hold for any $\Lambda$ and $\Lambda^a$, which is indeed the case.
Then instead of Eqs.(\ref{88}),(\ref{89}) in which we have the integration
over $\xi$, we obtain the equations without the integration over $\xi$:
\be
    \left ( - \, {{\Lambda} \over {2 \kappa |f|}} \eta^{\mu \nu} {{{\DD}^2}
    \over {{\DD} X^{\mu} (\xi) {\DD} X^{\nu} (\xi)}} - {{\Lambda} \over {2}} -
    {\cal E} \right ) \phi = 0
\label{90} 
\ee

\be
     \p_a  X^{\mu} (\xi)\, \left ( {{\delta \ } \over 
     {\delta X^{\mu} (\xi)}} + 
     {1\over 2} \Gamma_{\mu(\xi) \nu(\xi')}^{\nu(\xi')} \right ) \phi = 0
\label{91}
\ee

\nnn The last equations are obtained from (\ref{88}),(\ref{89}) after
writing the energy as the integral of the energy density ${\cal E}$ over
the membrane, $E = \int {\dd}^n \xi \, \sqrt{|f|} \, {\cal E}$, taking into
account that $K = \int {\dd}^n \xi \, \sqrt{|f|} \, \kappa \Lambda$, and
omitting the integration over $\xi$. 

Equations (\ref{90}),(\ref{91}), with ${\cal E} = 0$,
are indeed the quantum analogs of the classical $p$-brane constraints
used in the literature \cite{1},\cite{7} and
their solution $\phi$ represent states of a conventional, constrained,
$p$-brane with the tension $\kappa$. When ${\cal E} \neq 0$ the
preceding statement still holds, provided that ${\cal E} (\xi) $ is
proportional to $\Lambda (\xi)$, so that the quantity 
$\kappa (1 - 2 {\cal E}/\Lambda)$
is a constant, identified with the effective tension. Only the particular
stationary states (as indicated above)
correspond to the conventional, Dirac-Nambu-Goto $p$-brane states, but in
general they correspond to a sort of the wiggly membranes \cite{6,4a,5a}.

\vs{1cm}

{\bf 4.4. Dimensional reduction of the Schr\" odinger equation}

\vs{5mm}

Let us now consider the {\it Case 1}.
Our membrane has signature (+ - - - ... ) and is actually an
$n$-dimensional worldsheet. The index $a$ of the worldsheet coordinates
$\xi^a$ assumes the values $a = 0,1,2,...,p$, where $p = n - 1$.

Among all possible wave functional
satisfying Eqs.(\ref{64c}) there are also the special ones
for which it holds (for an example see see Eqs.(\ref{A50a}),
(\ref{117})-(\ref{119}))
\be
     {{\delta \psi} \over {\delta X^{\mu} (\xi^0, \xi^i)}} = 
     \delta (\xi^0 - \xi_{\Sigma}^0 ) (\p_0 X^{\mu} \p_0 X_{\mu})^{1/2}
     {{\delta \psi} \over {\delta X^{\mu} (\xi_{\Sigma}^0, \xi^i)}}
     \quad , \quad \quad i = 1,2,...,p=n-1
\label{98}
\ee

\nnn where $\xi_{\Sigma}^0$ is a fixed value of the time like coordinate
$\xi^0$. In the compact tensorial notation in membrane's space
${\cal M}$ Eq.(\ref{98}) reads
\be
     \p_{\mu (\xi^0,\xi^i)} \phi = \delta (\xi^0 - \xi_{\Sigma}^0) \, 
     (\p_0 X^{\mu} \p_0 X_{\mu})^{1/2} 
     \p_{\mu (\xi_{\Sigma}^0, \xi^i)} \psi
\label{99}
\ee

\nnn Using (\ref{99}) we find that dimension of the Laplace operator in
${\cal M}$ is lowered:
    $${\DD}^{\mu (\xi)}{\DD}_{\mu (\xi)} \psi = \int {\dd}^n \xi \,
    {{\Lambda} \over {\kappa \sqrt{|f|}}} \, \eta^{\mu \nu} {{{\DD}^2 \psi}
    \over
    {{\DD} X^{\mu} (\xi) {\DD} X^{\nu} (\xi)}}$$
     $$= \int {\dd} \xi^0 \, {\dd}^p \xi
    {{\Lambda} \over {\kappa \sqrt{|f|}}} (\p_0 X^{\mu} \p_0 X_{\mu})^{1/2}
    \, \eta^{\mu \nu}
    \delta (\xi^0 - \xi_{\Sigma}^0){{{\DD}^2 \psi} \over
    {{\DD} X^{\mu} (\xi^0, \xi^i) {\DD} X^{\nu} (\xi_{\Sigma}^0, \xi^i)}}$$
\be
     =
    \int {\dd}^p \xi
    {{\Lambda} \over {\kappa \sqrt{|{\bar f}|}}} \eta^{\mu \nu}
    {{{\DD}^2 \psi} \over
    {{\DD} X^{\mu} (\xi_{\Sigma}^0, \xi^i) X^{\nu} (\xi_{\Sigma}^0, \xi^i)}} =
    \int {\dd}^p \xi
    {{\Lambda} \over {\kappa \sqrt{|{\bar f}|}}} \eta^{\mu \nu} {{{\DD}^2 \psi} \over
    {{\DD} X^{\mu} (\xi^i) {\DD} X^{\nu} (\xi^i)}}
\label{100}
\ee

\nnn Here ${\bar f} \equiv \mbox{\rm det} {\bar f}_{ij}$ is
the determinant of the
induced metric ${\bar f}_{ij} \equiv \p_i X^{\mu} \p_j X_{\mu}$ on $V_p$,
and it is related to the determinant $f \equiv \mbox{\rm det} f_{ab}$
of the induced metric $f_{ab} = \p_a X^{\mu} \p_b X_{\mu}$ on $V_{p+1}$
according to $f = {\bar f} \, \p_0 X^{\mu} \p_0 X_{\mu}$ (see refs. \cite{8},
\cite{4a}-\cite{4}).

The differential
operator in the last expression of Eq.(\ref{100}) (where we identified
$X^{\mu (\xi_{\Sigma}^0, \xi^i)} \equiv X^{\mu (\xi^i)}$) acts in the space of
$p$-dimensional membranes, though the original ope\-rator we started from
acted in the space of $(p+1)$-dimensional membranes ($n=p+1$). This comes
from the fact that our special functional, satisfying (\ref{98}), has
vanishing functional derivative $\delta \psi/ \delta X^{\mu} (\xi^0, \xi^i)$
for all values of $\xi^0$, except for $\xi^0 = \xi_{\Sigma}^0$. Expression
(\ref{98}) has its finite dimensional analog in the relation
$\p \phi /\p x^A = {\delta_A}^{\mu} \, \p \phi / \p x^\mu$,
$A = 0,1,2,3,...,3 + m$, $\mu = 0,1,2,3$, which says that the field $\phi (
x^A)$ is constant along the extra dimensions. For such a field the
($4+m$)-dimensional Laplace expression $\eta^{AB} {{\p^2 \phi} \over
{\p x^A \p x^B}}$ reduced to the 4-dimensional expression
$\eta^{\mu \nu} {{\p^2 \phi} \over {\p x^\mu \p x^\nu}}$.

The above procedure can be performed not only for $\xi^0$, but for any of
the coordinates $\xi^a$; it applies both to {\it Case 1} and {\it Case 2}.

Using (\ref{99}),(\ref{100}) we thus find that for such a special
wave functional $\psi$ the equation (\ref{64c}),
which describes a state of a $(p+1)$-dimensional membrane,
reduces to the equation for a $p$-dimensional membrane. This an
important finding. Namely, at the beginning we may assume a certain dimension
of a membrane and then consider lower dimensional membranes as
particular solutions to the higher dimensional equation. This means
that the point particle theory (0-brane), the string theory (1-brane), and
in general a $p$-brane theory for arbitrary $p$ are all contained in
the theory of a $(p+1)$-brane.

\vs{1cm}

{\bf 4.5 A particular solution to the covariant Schr\" odinger equation}

\vs{5mm}

Let us now consider the covariant functional Schr\" odinger equation
(\ref{64c}) with the Hamiltonian operator (\ref{64c1}). The quantities
$\Lambda^a$ are arbitrary in principle. For simplicity we take now
$\Lambda^a = 0$. Additionally we also take a $\tau$-independent
$\Lambda$, so that ${\dot \rho} = 0$. Then Eq.(\ref{64c}) becomes
simply ($\hbar = 1$)
\be
   i \, {{\p \psi} \over {\p \tau}} = -\, {1 \over 2} \,
   ({\DD}^{\mu (\xi)} \DD_{\mu (\xi)} + K ) \psi
\label{101}
\ee

\nnn The operator on the right hand side is the infinite dimensional
analog of the covariant Klein-Gordon operator. Using the definition of the
covariant derivative (\ref{A17}) and the corresponding affinity  we have
\cite{4}
\be    
    {\DD}_{\mu (\xi)} {\DD}^{\mu (\xi)} \psi
    =  \rho^{\mu (\xi) \nu (\xi')}
   {\DD}_{\mu (\xi)}{\DD}_{\nu (\xi')} \psi
  =  \rho^{\mu (\xi) \nu (\xi')} \left (\p_{\mu(\xi)} \p_{\nu(\xi')} \psi -
     \Gamma_{\mu(\xi) \nu(\xi')}^{\alpha(\xi'')} \, \p_{\alpha(\xi'')}
     \psi \right )
\label{102} 
\ee

\nnn The affinity is explicitly
\be
      \Gamma_{\mu(\xi) \nu(\xi')}^{\alpha(\xi'')} = {1 \over 2}
      \rho^{\alpha(\xi'') \beta( \xi''')}\,
      \left ( \rho_{\beta(\xi''') \mu(\xi),\nu(\xi')} +
      \rho_{\beta(\xi''') \nu(\xi'), \mu(\xi)} - \rho_{\mu(\xi) \nu(\xi')
      , \beta(\xi''')} \right )
\label{102a}
\ee
      
\nnn where the metric is given by (\ref{A9}),(\ref{A13}) with
$\alpha = \kappa/\Lambda$ and $g_{\mu \nu} = \eta_{\mu \nu}$.
Using
\be
     \rho_{\beta(\xi'') \mu(\xi),\nu(\xi')} = \eta_{\mu \nu}\,
     \alpha (\xi) \delta(\xi - \xi'') {{\delta \sqrt{|f(\xi)|}} \over
     {\delta X^{\nu} (\xi')}} = \eta_{\mu \nu}\,
     \alpha (\xi) \delta(\xi - \xi'') \sqrt{|f(\xi)|} \p^a X_{\nu} (\xi)
     \p_a \delta (\xi - \xi')
\label{102b}
\ee     
     
\nnn the equation (\ref{102}) becomes
      $${\DD}_{\mu (\xi)} {\DD}^{\mu (\xi)} \psi = \rho^{\mu (\xi) \nu (\xi')}
   {{\p^2 \psi} \over {\p X^{\mu (\xi)} X^{\nu (\xi')}}}
     - {{\delta (0)} \over {\kappa}} \int {\dd}^n \xi \, 
    {{\delta \psi} \over {\delta X^{\mu} (\xi)}}$$
\be   
   \times \; \left [ {N\over 2}
    {{\Lambda} \over {\sqrt{|f|}}} \, {1\over {\sqrt{|f|}}} \,
    {\p}_{a} (\sqrt{|f(\xi)|} \, \, {\p}^{a} X^{\mu})    
    + ({N\over 2} + 1)
    \Lambda {\p}^{a} X^{\mu}\, \p_a ({1\over {\sqrt{|f(\xi)|}}}) 
    +{{{\p}^{a} X^{\mu} \p_a \Lambda} \over {\sqrt{|f(\xi)|}}} \right ]
\label{107}
\ee

\nnn where $N = \eta^{\mu \nu} \eta_{\mu \nu}$ is dimension of spacetime.
In deriving Eq.(\ref{107}) we encountered the expression $\delta^2
(\xi - \xi')$ which we replaced by the corresponding approximate expression
$F(a, \xi-\xi') \delta(\xi - \xi')$ where$F(a, \xi-\xi')$ is any finite
function, e.g.$(1/\sqrt{\pi} a) {\rm exp} [-(\xi-\xi')2/a^2]$, which in the
limit $a \rightarrow 0$ becomes $\delta(\xi-\xi')$. The latter limit was
taken after performing all the integrations, and $\delta(0)$should be
considered as an abbreviation for $\lim_{a\to 0} F(a,0)$.

The infinity $\delta(0)$ in an expression such as (\ref{107}) can be
regularized by taking into account the plausible assumption that a
generic physical object is actually a fractal (i.e., an object with 
detailed structure on all scales). The objects $X^{\mu} (\xi)$ which we are
using in our calculations are well behaved functions with definite
derivatives and should be considered as approximations to the actual
physical objects. This means that a description with a given $X^{\mu} (\xi)$
bellow a certain scale $a$ has no physical meaning. In order to make a
physical sense of the expression (\ref{107}), $\delta (0)$ should
therefore be replaced by $F(a,0)$. Choice of the scale $a$ is arbitrary
and determines the precision of our description. (C.f., the length of a coast
depends on the scale of the units which it is measured with.)

Expression (\ref{102}) is analogous to the corresponding finite
dimensional expression. In analogy to a finite dimensional case,
a metric tensor ${\rho '}_{\mu(\xi)\nu(\xi')}$ obtained from
${\rho }_{\mu(\xi)\nu(\xi')}$ by a coordinate transformation (\ref{A5})
belongs to the same space ${\cal M}$  and is equivalent to
${\rho }_{\mu(\xi)\nu(\xi')}$. Instead of a finite number of
coordinate conditions which
express a choice of coordinates, we have now infinite coordinate
conditions. The second term in eq.(\ref{107}) becomes zero if we take
\be
  {N\over 2} {{\Lambda} \over {\sqrt{|f|}}} \, {1\over {\sqrt{|f|}}} \,
    {\p}_{a} \left (\sqrt{|f(\xi)|} \, \, {\p}^{a} X_{\mu} \right ) 
    + \left (
    ({N\over 2} + 1) \Lambda \p_a ({1\over {\sqrt{|f(\xi)|}}}) +
    {{\p_a \Lambda} \over {\sqrt{|f(\xi)|}}} \right ) {\p}^{a} X_{\mu} = 0
\label{DD1} 
\ee

\nnn and these are just possible coordinate conditions in the
membrane space ${\cal M}$. Eq.(\ref{DD1}) together with boundary
conditions determines a family of functions $X^{\mu} (\xi)$   
for which the functional $\psi$ is defined; in the operator theory such
a family is called the domain of an operator. Choice of a family
of functions $X^{\mu} (\xi)$ is in fact choice of  coordinates (a gauge)
in ${\cal M}$.

If we contract Eq.(\ref{DD1}) by $\p_b X_{\mu}$ and take into account
the identity (\ref{8}) we find
\be
    ({N\over 2} + 1) \Lambda \p_a ({1\over {\sqrt{|f(\xi)|}}}) +
    {{\p_a \Lambda} \over {\sqrt{|f(\xi)|}}} = 0
\label{DD2}
\ee

\nnn From (\ref{DD2}) and (\ref{DD1}) we have
\be
   {1 \over { \sqrt{|f(\xi)|}}} \,
 {\p}_{a} \left ( \sqrt{|f(\xi)|} \, \, {\p}^{a} X_{\mu} \right) = 0
\label{DD3}
\ee

\nnn  Interestingly, the gauge condition (\ref{DD1}) in ${\cal M}$
automatically implies the gauge condition (\ref{DD2}) in $V_n$. The
latter condition is much simplified if we take $\Lambda \ne 0$
satisfying $\p_a \Lambda = 0$; then for $|f| \ne 0$ Eq.(\ref{DD2}) becomes
\be
                        \p_a \sqrt{|f|} = 0
\label{DD4}
\ee

\nnn which is just the gauge condition considered by Schild \cite{Schild}
and Eguchi \cite{Eguchi}.
    
In the presence of the condition (\ref{DD1}) (which is just a gauge
or coordinate condition in the function space ${\cal M}$) the
functional Schr\" odinger equation (\ref{101}) can be written in the
form
\be
          i \, {{\p \psi} \over {\p \tau}} = - {1 \over 2} \int {\dd}^n \xi
     \left ( {{\Lambda} \over {\sqrt{|f|} \kappa}} \, {{{\delta}^2}
    \over {\delta X^{\mu} (\xi) \delta X_{\mu} (\xi)}} + \sqrt{|f|} \,
    \Lambda \kappa \right ) \psi
\label{109}
\ee

\nnn A particular solution to Eq.(\ref{109}) can be obtained by considering the
following eigenfunctions of the momentum operator
\be
     {\hat p}_{\mu (\xi)} \psi_p [X^{\mu (\xi)}] = 
     p_{\mu (\xi)} \psi_p [X^{\mu (\xi)}]
\label{110}
\ee

\nnn with
\be    
    \psi_p [X^{\mu (\xi)}] = {\cal N} \mbox{\rm exp} \left [i \int_{X_0}^X
    p_{\mu (\xi)} \, {\dd} {X'}^{\mu (\xi)} \right ]
\label{111a}
\ee

\nnn This last expression is invariant under reparametrizations of $\xi^a$
(Eq.(\ref{A2})) and of $X^{\mu} (\xi^a)$ (Eq.(\ref{A5})). The momentum
field $p_{\mu (\xi)}$ in general functionally depends on $X^{\mu (\xi)}$
and satisfies  Eqs.(\ref{43a5}),(\ref{43a5a}). In particular $p_{\mu (\xi)}$
may be just a constant field, such that $\p_{\nu (\xi')} p_{\mu (\xi)} = 0$.
Then (\ref{111a}) becomes
\be
   \psi_p [X^{\mu (\xi)}] = {\cal N} \mbox{\rm exp} \left [i \int 
   {\dd}^n \xi \,
    p_{\mu} (\xi) (X^{\mu} (\xi) - X_0^{\mu} (\xi)) \right ] =
    {\cal N} \mbox{\rm exp} \left [i p_{\mu (\xi)} (X^{\mu (\xi)} -
    X_0^{\mu (\xi)}) \right ] 
\label{111}
\ee

\nnn The latter expression holds only in a particular parametrization (Eq.(\ref
{A5})) of ${\cal M}$ space, but it is still invariant with respect to
reparametrizations of $\xi^a$. 

Let the $\tau$-dependent wave functional be    
    $$\psi_p [\tau, X^{\mu} (\xi)] =$$
    $${\cal N} \mbox{\rm exp} \left [i \int {\dd}^n \xi \,
    p_{\mu} (\xi) \left ( X^{\mu} (\xi) - X_0^{\mu} (\xi) \right ) \right ]
    \mbox{\rm exp} \left [- {{i \tau} \over 2} \int \Lambda {\dd}^n \xi
    \left ( {{p^{\mu} (\xi) p_{\mu} (\xi)} \over {\sqrt{|f|} \kappa}} -
    \sqrt{|f|} \kappa \right ) \right ]$$
\be    
    \equiv {\cal N} \mbox{\rm exp} \left [ i 
     \int {\dd}^n \xi \, {\cal S} \right ]
\label{112}
\ee

\nnn where $f \equiv {\rm det} \, \p_a X^{\mu} (\xi) \p_b X_{\mu} (\xi)$ should
be considered as a functional of $X^{\mu} (\xi)$ and independent of $\tau$.
From (\ref{112}) we find
   $$- \, i \, {{\delta \psi_p [\tau,X^{\mu}]} \over {\delta X^{\alpha}}}
   = - \, i \, \left ( {{\p {\cal S}} \over {\p X^{\alpha}}} - \p_a
   {{\p {\cal S}} \over {\p \p_a X^{\alpha}}} \right ) 
   \psi_p [\tau, X^{\mu}]$$

\be
   = p_{\alpha} \psi_p
    - \, \tau \p_a (\sqrt{|f|} \, \p^a X_{\mu} ) {{\Lambda} \over {2 \kappa}}
   \, \left ( {{p^2} \over {|f|}} + \kappa^2 \right ) \psi_p - \tau \, 
   \sqrt{|f|}
   \, \p^a X_{\mu} \p_a \left [ {{\Lambda} \over {2 \kappa}}
   \, \left ( {{p^2} \over {|f|}} + \kappa^2 \right ) \right ] \psi_p
\label{113}
\ee

\nnn Let us now take the gauge condition (\ref{DD3}). Additionally, let us
assume
\be
    \p_a \left [ {{\Lambda} \over {2 \kappa}}
   \, \left ( {{p^2} \over {|f|}} + \kappa^2 \right ) \right ] \equiv
   \kappa \, \p_a \mu = 0
\label{114}
\ee

\nnn By inspecting the classical equations of motion (\ref{5}) with
$\Lambda^a = 0$ and Eq.(\ref{9}) we see that Eq.(\ref{114}) is satisfied
when the momentum of a classical membrane does not change with $\tau$, i.e.
\be
    { {{\dd} p_{\mu}} \over {{\dd} \tau} } = 0
\label{115}
\ee

\nnn Then the membrane satisfies
the minimal surface equation (\ref{12}) which is just our gauge condition
(\ref{DD3}) in the membrane space ${\cal M}$. \footnote
  {The reverse is not necessarily true: the imposition of the gauge
  condition (\ref{DD3}) does not imply (\ref{114}),(\ref{115}). The latter
  are additional assumptions which fix a possible congruence of
  trajectories (i..e. of $X^{\mu (\xi)} (\tau)$) over which the wave
  functional is defined.}
When ${\dot \Lambda} = 0$,
 the energy $E \equiv \int \Lambda {\dd}^n \xi
    \left ( {{p^{\mu} p_{\mu}} \over {\sqrt{|f|} \kappa}} -
    \sqrt{|f|} \kappa \right )$ is a constant of motion. Energy conservation
in the presence of Eq.(\ref{115}) implies
\be
    {{{\dd} \sqrt{|f|}}\over {{\dd} \tau}} = 0
\label{115a}
\ee
        
Our stationary state (\ref{112}) is thus defined over a congruence of
classical trajectories satisfying (\ref{12}) and (\ref{115}) which imply also
(\ref{114}) and (\ref{115a}). Eq.(\ref{113}) then becomes simply
\be
     - \, i \, {{\delta \psi_p} \over {\delta X^{\alpha}}} =
       p_{\alpha} \psi_p 
\label{116}
\ee

\nnn Using (\ref{116}) it is straightforward to verify that (\ref{112})
is a particular solution to the Schr\" odinger eqation (\ref{101}). In
Ref.\cite{4} we found the same relation (\ref{116}), but by using a
different, more involved procedure.

\vs{1cm}

{\bf 4.6 The wave packet}

\vs{5mm}

From the particular solutions (\ref{112}) we can compose a wave packet
\be
      \psi [\tau, X^{\mu} (\xi)] = \int {\cal D} p \, c[p] \,
      \psi_p [\tau, X^{\mu} (\xi)]
\label{116a}
\ee

\nnn which is also a solution to the Schr\" odinger equation (\ref{101}).
However, since  all $\psi_p [\tau, X^{\mu} (\xi)]$ entering (\ref{116a})
are supposed to belong
to a restricted class of particular solutions with $p_{\mu} (\xi)$
which does not functionally depend on $X^{\mu} (\xi)$, a wave packet 
of the form (\ref{116a}) cannot represent every possible membrane's state.
This is just a particular kind of wave packet; a general wave packet can be
formed from a complete set of particular solutions which are not restricted to
momenta $p_{\mu (\xi)}$ satisfying $\p_{\nu (\xi')} p_{\mu (\xi)} = 0$,
but allow for $p_{\mu (\xi)}$ which do depend on $X^{\mu (\xi)}$.
Treatment of such a general case is beyond the scope of the present paper.
Here we shall try to demonstrate some illustrative properties of the wave
packet (\ref{116a}). 

In the definition of the invariant measure in
momentum space we use the metric (\ref{A13}) with $\alpha = \kappa/\Lambda$:
\be
    {\cal D} p \equiv \prod_{\xi ,\mu} \mbox{$ {\left ( 
   {{\Lambda} \over {\sqrt{|f|} \, \kappa }} \right ) }^{1/2} $}
   {\dd } p_{\mu} (\xi) 
\label{A48}
\ee

\nnn Let us take\footnote{
This can be written compactly as $c[p] = {\cal B} \,\mbox{\rm exp}
\left [ -{1 \over 2} \rho_{\mu (\xi) \nu (\xi'')}
(p^{\mu (\xi)} - p_0^{\mu (\xi)}) (p^{\nu (\xi')} - p_0^{\nu (\xi')})
{\sigma_{(\xi')}}^{(\xi'')} \right ]$, where ${\sigma_{(\xi')}}^{(\xi'')} =
\sigma (\xi') \delta(\xi', \xi'')$. Since the covariant derivative of
the metric is zero, we have that ${\DD}_{\mu (\xi)} c[p] = 0$.
Similarly, the measure ${\cal D} p = (\mbox{\rm Det} \,
\rho_{\mu (\xi) \nu (\xi')})^{1/2} \prod_{\mu , \xi} {\dd} p_{\mu} (\xi)$,
and the covariant derivative of the determinant is zero. Therefore
${\DD}_{\mu (\xi)} \int {\cal D} p \, c[p] \, \psi_p [\tau, X^{\mu} (\xi)] =
\int {\cal D} p \, c[p] \, {\DD}_{\mu (\xi)}\psi_p [\tau, X^{\mu} (\xi)]$.
This confirms that the superposition (\ref{116a}) is a solution if
$\psi_p$ is a solution of (\ref{64c}).}

\be
    c[p] = {\cal B} \, \mbox{\rm exp} \left [ -{1 \over 2} \int d^n \xi \,
    {{\Lambda} \over {\sqrt{|f|} \kappa}} (p^{\mu} - p_0^{\mu})^2 \,
     \sigma (\xi) \right ]
\label{A49}
\ee

\nnn where ${\cal B} = \lim_{\Delta \xi \to 0} \prod_{\xi,\mu}
{\left ( {{\Delta \xi \, \sigma (\xi)} \over {\pi}} \right ) }^{1/4}$ is the
normalization constant, such that $\int {\cal D} p \, c^{*} [p] c[p] = 1$.
For the
normalization constant ${\cal N}$ occurring in (\ref{112}) we take
${\cal N} = \lim_{\Delta \xi \to 0} \, \prod_{\xi,\mu} {\left ( 
{{\Delta \xi} \over {2 \pi}} \right )}^{1/2}$.
From (\ref{116a})-(\ref{A49}) and (\ref{112}) we have
$$
   \psi[\tau, X (\xi)] = 
  \lim_{\Delta \xi \to 0} \prod_{\xi,\mu} \int
   \mbox{$
   {\left ( {{\Delta \xi } \over {2 \pi}} \right )}^{1/2}
   {\left ( {{\Delta \xi \, 
   \sigma (\xi)} \over {\pi}} \right )}^{1/4}
   { \left ({{\Lambda} \over {\sqrt{|f|} \kappa}}\right )}^{1/2} $}
    \, \dd p_{\mu} (\xi)
$$    

$$
    \times \mbox{\rm exp} \left [- {{\Delta \xi} \over 2}
   {{\Lambda} \over {\sqrt{|f|} \kappa}}
   \left ( (p^{\mu} - p_0^{\mu})^2 \sigma (\xi)
   - 2 i {{\sqrt{|f|} \kappa} \over {\Lambda}} p_{\mu}
   (X^{\mu} - X_0^{\mu}) + i \tau p_{\mu} p^{\mu} \right ) \right ]$$

\be
   \times \mbox{\rm exp} \left [ {{i \tau} \over 2} \int 
  {\dd}^n \xi \, \sqrt{|f|} \Lambda \kappa \right ]
\label{A54}
\ee
 
\nnn We assume no summation over $\mu$ in the exponent of the above
expression and no integration (actually summation) over $\xi$,
because these operations are now already included in the product which
acts on the whole expression.
Because of the factor $
{ \left ( {{\Delta \xi \Lambda} \over {\sqrt{|f|} \kappa}} \right ) }^{1/2}$
occurring in the measure and the same factor in the exponent, the integration
over $p$ in Eq.(\ref{A54}) can be performed straightforwardly. The result is
   $$\psi[\tau , X] = 
   \mbox{$ \left [ \lim_{\Delta \xi \to 0} {\prod}_{\xi,\mu}
   {\left ( {{\Delta \xi \, \sigma } \over {\pi}} \right )}^{1/4}
    {\left ( {1 \over {\sigma + i \tau }} \right ) }^{1/2} \right ] $}$$

\be    
  \times \, \mbox{\rm exp} \left [ \int d^n \xi
   {{\Lambda} \over {\sqrt{|f|} \kappa}}
 \left (
 {{\left ( i {{\sqrt{|f|} \kappa} \over {\Lambda}} 
  (X^{\mu} - X_0^{\mu}) + p_0^{\mu} \sigma \right )^2} \over
  {2(\sigma + i \tau)}} - {{p_0^2 \sigma} \over 2} \right ) \right ]
  \mbox{\rm exp} \left[
   {{i \tau} \over 2} \int {\dd}^n \xi \, \sqrt{|f|} \Lambda \kappa \right ] 
\label{A50}
\ee

\nnn Eq.(\ref{A50}) is a generalization of the familiar Gaussian wave packet.
At $\tau = 0$ Eq.(\ref{A50}) becomes
$$
      \psi[0, X] =  \mbox{$ \left [ \lim_{\Delta \xi \to 0} {\prod}_{\xi,\mu}
   {\left ( {{\Delta \xi} \over {\pi \sigma}} \right )}^{1/4}
   \right ] $}
   \mbox{\rm exp} \left [ - \int \dd^n \xi {{\sqrt{|f|} \kappa} \over {\Lambda}}
   \, {{\left ( X^{\mu} (\xi) - X_0^{\mu} (\xi) \right )^2} \over 
   {2 \sigma (\xi)}} \right ]$$
\be
      \times \mbox{\rm exp} \left [i \int p_{0 \mu} (X^{\mu} - X_0^{\mu}) \,
      {\dd}^n \xi \right ]
\label{A50a}
\ee

The probability density is given by
   $$|\psi [\tau , X ] |^2  = $$
\be   
   \mbox{$ \left [ \lim_{\Delta \xi \to 0} \prod_{\xi,\mu}
   {\left ( {{\Delta \xi \, \sigma } \over {\pi}} \right )}^{1/2}
    {\left ( {1 \over {{\sigma}^2 + {\tau }^2}}\right ) }^{1/2} \right ] $}
    \mbox{\rm exp} \left [ {- \int d^n \xi
     {{\sqrt{|f|} \kappa} \over {\Lambda}}
   {{( X^{\mu} - X_0^{\mu}  -  {{\Lambda} \over {\sqrt{|f|} \kappa}}
    p_0^{\mu} \tau)^2} \over {({\sigma}^2  + {\tau }^2)/ \sigma}}} \right ]
\label{A51}
\ee

\nnn and the normalization constant, though containing the infinitesimal
$\Delta \xi$, gives precisely $\int |\psi|^2 {\cal D} X = 1$.

From (\ref{A51})
we find that the motion of the centroid membrane of our particular wave
packet is determined by the equation
\be
   X_{\mbox{\rm c}}^{\mu} (\tau, \xi) = X_0^{\mu} (\xi) + 
   {{\Lambda} \over {\sqrt{|f|} \kappa}} \, p_0^{\mu} (\xi) \tau
\label{A52}
\ee

\nnn  From the classical equation of motion (\ref{5}) (see also (\ref{115})),
(\ref{115a}) we indeed obtain a solution of the form (\ref{A52}).
At this point it is interesting to observe that the classical null
strings considered,
within different theoretical frameworks, by Schild \cite{Schild} and
Roshchupkin et al. \cite{Zheltukhin} also move according to the
equation (\ref{A52}).

Function $\sigma (\xi)$ in Eqs.(\ref{A49})-(\ref{A51})
is arbitrary; choice of $\sigma (\xi)$
determines how the wave packet is prepared. In particular, we may
consider {\it Case 1} of Sec. 2.2 and take
$\sigma (\xi)$ such that the wave packet of a $p+1$-dimensional
membrane ${\cal V}_{p+1}$ 
is peaked around a space-like $p$-dimensional membrane ${\cal V}_p$.
This means that the wave functional localizes ${\cal V}_{p+1}$ much more
sharply around ${\cal V}_p$ than in other regions of spacetime.
Effectively, such a wave packet describes the $\tau$-evolution of
${\cal V}_p$ (though formally it describes the $\tau$-evolution of
${\cal V}_{p+1}$). This can be clearly seen by taking the following
limiting form of the wave packet (\ref{A50a}), such that
\be
    {1 \over {\sigma (\xi)}} = {{\delta (\xi^0 - \xi_{\Sigma}^0)} \over
    {\sigma (\xi^i) (\p_0 X^{\mu} \p_0 X_{\mu})^{1/2} }} \; \; ,
    \quad \quad i = 1,2,...,p
\label{117}
\ee

\nnn and choosing
\be
     p_{0 \mu} (\xi^a) = {\bar p}_{0 \mu} (\xi^i) \delta (\xi^0 -
     \xi_{\Sigma}^0)
\label{118}
\ee

\nnn Then the integration over $\delta$-function gives in the exponent
of Eq.(\ref{A50a}) the expression
\be
     \int \dd^p \xi {{\sqrt{|{\bar f}|} \kappa} \over {\Lambda}}
      \, (X^{\mu} (\xi^i) - X_0^{\mu} (\xi^i)
     )^2 {1 \over {2 \sigma (\xi^i)}} + i \int {\dd}^p \xi \, {\bar p}_{0 \mu}
     (X^{\mu} (\xi^i) (\xi^i) - X_0^{\mu} (\xi^i) )
\label{119}
\ee

\nnn so that Eq.(\ref{A50a}) becomes a wave functional of a $p$-dimensional membrane
$X^{\mu} (\xi^i)$. Here again ${\bar f}$ is the determinant of the induced
metric on $V_p$, while $f$ is the determinant of the induced metric on
$V_{p+1}$. One can verify that such a wave functional (\ref{A50a}),
(\ref{119}) satisfies the relation (\ref{98}).

The analogous considerations to those described above hold for the
{\it Case 2} as well,
so that a wave functional of a $(p-1)$-brane can be considered as a
limiting case of a $p$-brane's wave functional.

\vs{1cm}

{\bf 5. Conclusion}

\vs{5mm}

We started to elaborate a theory of relativistic $p$-branes which is more
general than the theory of conventional, constrained, $p$-branes. In
the proposed generalized theory, $p$-branes are unconstrained, but among
the solutions to the classical and quantum equations of motion there
are also the usual, constrained, $p$-branes. A strong motivation for such
a generalized approach is the elimination of the well known difficulties
due to the presence of constraints. Since the $p$-brane theories are
still at the stage of development and have not yet been fully confronted
with observations, it makes sense to consider an enlarged set of
classical and quantum $p$-brane states, such as e.g. proposed in the
present and some previous works \cite{4a}-\cite{4}.
What we gain is a theory without
constraints, still fully relativistic, which is straightforward both at
the classical and the quantum level, and is not in conflict with the
conventional $p$-brane theory.

Our approach might shed more light on some very interesting
development concerning the duality \cite{3}, such as one of
strings and five-branes,
and the interesting interlink between $p$-branes of various dimensions
$p$. In the present paper we have demonstrated how a higher dimensional
$p$-brane equation naturally contains lower dimensional $p$-branes as
solutions.

The highly non trivial concept of unconstrained membranes enables us
to develop the elegant formulation of "point particle" dynamics in the
infinite dimensional space ${\cal M}$. It is fascinating that the action,
canonical and Hamilton formalism and, after quantization, the
Schr\" odinger equation all look as nearly trivial extensions of the
correspondings objects in the elegant Fock-Stueckelberg-Schwinger-DeWitt
proper time formalism for a point particle in curved space.
Just this "triviality", or better simplicity is
a distinguished feature of our approach and we have reasons to
expect that also the $p$-brane gauge field theory - not yet a
completely solved problem - can be straightforwardly formulated along
the lines indicated in the present paper.

\end{document}